\documentclass[3p]{elsarticle}
\pdfoutput=1

\journal{Ultramicroscopy}

\usepackage{amsmath}
\usepackage{bm}
\usepackage{color}
\usepackage{xfrac}
\usepackage{natbib}
\usepackage{amssymb}
\usepackage{lineno,hyperref}
\usepackage{float}

\begin{document}
\begin{frontmatter}

\title{Thon rings from amorphous ice and implications of beam-induced 
Brownian motion in single particle electron cryo-microscopy}

\author[MRC]{G.~McMullan \corref{cor1}}
\ead{gm2@mrc-lmb.cam.ac.uk}
\cortext[cor1]{Corresponding author}
 \author[MRC]{K.R.~Vinothkumar}
 \author[MRC]{R. Henderson}
 \address[MRC]{MRC Laboratory of Molecular Biology, Francis Crick Avenue, Cambridge, CB2 0QH, U.K.}

\begin{abstract}
We have recorded dose-fractionated electron cryo-microscope images of thin films of
pure flash-frozen amorphous ice and pre-irradiated amorphous 
carbon on a Falcon~II direct electron detector
using 300 keV electrons. 
We observe Thon rings \cite{Thon1966} 
in both the power spectrum of the summed frames and the sum of power spectra 
from the individual frames. 
The Thon rings from amorphous carbon images are always more visible in the
power spectrum of the summed frames whereas those of amorphous ice are more visible in the 
sum of power spectra from the individual frames.
This difference indicates that while pre-irradiated carbon 
behaves like a solid during the exposure, amorphous ice behaves like
a fluid with the individual water molecules undergoing beam-induced motion.
Using the measured  variation in the power spectra amplitude
with number of electrons per image we deduce that water molecules are 
randomly displaced by mean squared
distance of $\sim$ 1.1 \AA$^{2}$ for every incident 300~keV~e$^{-}$/\AA$^2$.
The induced motion leads to an optimal  
exposure with 300 keV electrons of 4.0~e$^{-}$/\AA$^2$ per image with which to see 
Thon rings centred around the strong 
3.7\AA\/ scattering peak from amorphous ice. 
The beam-induced movement of the water molecules generates
pseudo-Brownian motion of embedded macromolecules. The 
resulting blurring of single particle images contributes an additional
term, on top of that from radiation damage, to the minimum 
achievable B-factor for macromolecular structure determination.
\end{abstract}

\end{frontmatter}

\twocolumn
\section{Introduction}
During the 1980s, Dubochet and his colleagues \citep{dubochet_electron_1982,Dubochet1988}
developed a method for carrying out electron cryo-microscopy (cryoEM) of 
biological structures embedded in thin films of amorphous ice.  Their early
work involved careful comparison of the conditions needed to obtain thin 
films of amorphous rather than hexagonal or cubic crystalline ice \cite{dubochet_electron_1982}.
The method they developed consisted of rapidly freezing a thin film of 
water, or buffer solution, containing the biological structure of interest 
by plunging it into liquid ethane at a temperature just above that of liquid nitrogen.  
Images of the resulting specimen recorded using a suitable microscope could then 
be analysed to determine the structure. The method has since grown immensely 
in its power and popularity as improvements in technology 
have transformed its capability \cite{Kuhlbrandt28032014}. 

There has long been a debate within the electron cryo-microscopy community
as to whether Thon rings \cite{Thon1966} can or should be observed in power spectra
of images of pure plunge-frozen amorphous ice.  
Thon rings arise naturally in bright-field 
phase contrast images of any amorphous material and 
their origin is described in most standard texts
\cite{REIMER_2008,spence_high-resolution_2013}. 
The spacing and shape of the rings depend on the electron-optical parameters 
that describe the image, principally the amount of defocus and astigmatism.  
Thon rings from amorphous carbon are routinely used to adjust astigmatism and 
set the defocus of a microscope. Amorphous ice has approximately one half the density of 
amorphous carbon but as oxygen atoms scatter more strongly than carbon atoms \cite{REIMER_2008}
one might naively expect to see Thon rings of similar strength from 
amorphous ice and amorphous carbon.

In this paper we show that with the recently introduced CMOS direct electron detectors 
it is possible to see unambiguously Thon rings in images 
of high purity amorphous ice. The higher detective quantum efficiency, DQE, 
of the new detectors allows weaker signals to be seen and
the ability to collect images continuously in a dose fractionated (or movie) mode 
allows optimal exposure conditions to be chosen and new modes of image processing to be used, 
long after the sample has been removed
from the microscope.

\section{Theory}
\subsection{Definitions}
In this work the pixel value in an image at pixel $(r,s)$ is denoted $f_{r,s}$, and 
scaled to units of e$^{-}$/pixel.  The discrete Fourier transform, $f({\bm u})$, 
at spatial frequency $\bm u$ is 
calculated using 
\begin{equation}
f({\bm u})  = 
\sum_{r,s} f_{rs} \exp( -i2\pi {\bm u}{\cdot} {\bm x}_{rs}).
\label{eqn:FOURIER}
\end{equation}
and the power spectra, $S({\bm u})$, calculated using the convention
\begin{equation}
S({\bm u})  = \frac{1}{ N_\mathrm{tot}} | f({\bm u}) | ^2 = 
 \frac{1}{N_{\mathrm{tot}}} 
\bigg| \sum_{r,s} f_{rs} \exp( -i2\pi{\bm x}_{rs}.{\bm u}) \bigg|^2.
\label{eqn:POWER_SPECTRA} 
\end{equation}

\subsection{Noise whitened power spectra}
The modulation transfer function, MTF, and detective quantum efficiency, DQE, of
the Falcon~II detector are given in \cite{mcmullan_comparison_2014}. 
The MTF of the Falcon~II
drops rapidly with increasing spatial frequency corresponding to the Falcon~II
having a broad point spread function, PSF. For the
combination of pixel size and sensitive layer thickness used in the 
Falcon~II the PSF is chiefly determined by charge carrier diffusion in the sensitive layer 
rather than the scattering of incident electrons \cite{mcmullan_comparison_2014}. 
As a result the DQE of the Falcon~II does  not fall with the MTF.
The strong spatial frequency 
dependence of the MTF is however reflected in the power spectra from the Falcon~II and
can make it difficult to see weak signals. In this case it is advantageous to use 
the noise whitened power spectrum, $W({\bm u})$, 
given by 
\begin{equation}
W({\bm u})  =  S ({\bm u})/ \mathcal{N}({\bm u})
\label{eqn:WHITE_POWER_SPECTRA}
\end{equation}
in which $\mathcal{N}({\bm u})$ is the normalise noise power spectrum \cite{mcmullan_comparison_2014}.
For uniformly illuminated
images, noise whitening removes the spatial frequency dependence of $S({\bm u })$ by boosting
the signal at higher spatial frequencies. In particular it gives a
flat background on which small signals can more easily be seen.

In \cite{mcmullan_comparison_2014} it was shown that for the Falcon~II,
$\mathcal{N}({\bm u})$  could be estimated directly from the measured
modulation transfer function, MTF. 
Using $\mathcal{N}({\bm  u})$ estimated in this way results in 
only a few percent variation in 
$W({\bm u})$ over the range from near zero to the Nyquist frequency in images with 
no sample present.  In the present work it was found that the residual variation 
in $W({\bm u})$ with ${\bm u}$ could be further reduced by  using a fit
to the measured $S({\bm u})$ with model for
$\mathcal{N}({\bm  u})$ consisting of only a radially symmetric function and its low order aliased terms.
Note that  while $\mathcal{N}({\bm  u})$ is a two dimensional function almost all of 
the non-circularly symmetric components arise simply from the contributions of aliased terms.
Details of the fitting proceudure are given in the supplementary data.

\subsection{Probability distribution for values in the noise whitened power spectra}
\label{sec:hist}
For uniformly illuminated images the probability distribution for values of $W({\bm u})$ 
can be described using a single parameter.
With no sample, the real and imaginary components of $f({\bm u})$ 
in Eq.~\eqref{eqn:FOURIER} are independent Gaussian random variables \cite{rice_mathematical_1944}. 
As a result the squares of the real and imaginary components have $\chi^2$ distributions 
and their sum, $S({\bm u})$, an exponential distribution.  
Noise whitening sets the same average value, and hence exponential distribution for  
values of $W({\bm u})$, at all
spatial frequencies, $\bm u$. For any given spatial frequency, ${\bm u}$,  the probability 
that $W({\bm u})$ has value $y$ is therefore
\begin{equation}
\mathrm{Prob}\big[ y = W({\bm u})\big] = \exp( -y/\Gamma)/\Gamma
\label{eqn:PROB}
\end{equation}
in which $\Gamma$ is the mean value of $W({\bm u})$.

The DQE of the Falcon~II detector is principally limited by the
intrinsic variability of its response to individual electrons.
If the average gain and variance in the gain are 
$\bar g$ and   $\sigma_g^2$, respectively, then the value of noise power spectrum 
in the limit of zero spatial frequency with $n$ incident independent electrons per 
pixel is \cite{zweig_detective_1965}
\begin{equation}
S( 0)  =  \bar g^2 n  + \sigma_g^2  n
\label{eqn:SX}
\end{equation}
and
\begin{equation}
\mathrm{DQE}(0) = \bar g^2/(\bar g^2 + \sigma_g^2).
\label{eqn:DQE}
\end{equation} 
Using Eqns.~\eqref{eqn:WHITE_POWER_SPECTRA} to \eqref{eqn:DQE} gives
\begin{equation}
W(0) = W(u) = n/\mathrm{DQE}(0) = \Gamma
\label{eqn:WX}
\end{equation}
since the average value of $W({\bm u})$ is the same for all $\bm u$, 
by definition $\mathcal{N}( 0) = 1$ and
expressing the detector output in units of incident electrons 
sets the gain to unity.
The probability distribution for values of $W({\bm u})$ 
depends solely on $\Gamma=n /\mathrm{DQE}(0)$ via Eq.~\eqref{eqn:PROB}
and has both mean and standard deviation of $\Gamma$.
In particular, for the Falcon~II $\mathrm{DQE}(0)\sim 0.5 $ 
\cite{mcmullan_comparison_2014} and so 
$\Gamma  \sim 2 n $. 

\subsection{Power spectra of Dose fractionated series}
If a dose fractionated exposure consists  of 
$M$ frames with an average of $d$ uncorrelated electrons
per frame, from Eq.~\eqref{eqn:WX} the expected 
value of $W_{M,m}$ defined as
the sum of the $M/m$ noise whitened power spectra from the images
obtained by summing $m$ consecutive frames is 
\begin{equation}
W_{M,m} = \sum_i^{M/m}   (md/\mathrm{DQE}) =  \frac{D_{\mathrm{tot}}}{\mathrm{DQE}}.
\label{eqn:WM}
\end{equation}
in which $D_{\mathrm{tot}} = Md$ is the total number of electrons in the exposure.
Similarly the expected noise, $N_{M,m}$, in $W_{M,m}$ is
\begin{equation}
N_{M,m} = \sqrt{ \sum_i^{M/m}  (md/\mathrm{DQE})^2 }  = 
\frac{D_{\mathrm{tot}}}{\mathrm{DQE}}
\sqrt{\frac{m}{M}}.
\label{eqn:NOISE}
\end{equation}
While the value of $W_{M,m}$ is independent of $m$, the noise 
in sum, $N_{M,m}$,  grows as the square root of $m$. 
In practice residual correlation in pixel values between frames, 
such as from an offset drift, will result
in an $m$ dependence in $W_{M,n}$. 
The actual Falcon II detector used here had a 
small correlated shift in the row-reset between successive frames that results in
a 1\% drop in $W_{M,m}(u)$ between $m=1$ and $m=2$. However the 
largest systematic correlation comes from the 
applied per pixel gain correction of the images which results in an increase
in  $W_{M,m}(u)$ with increasing $m$ but with
careful gain calibration keeps the effect is less than 5\% in going from 
$m=1$ and $m=120$.

\subsection{Effect of random motion of water molecules on power spectra}
\label{sec:GAUSS}
At any given instant there is typically only a single high energy electron 
interacting with the sample during a typical cryoEM exposure.   An 
incident electron passing through a thin layer of amorphous ice will see
a particular conformation of atoms but the radiation damage resulting
from the interaction of the high energy electron with the amorphous ice
will perturb the conformation of atoms in the ice \cite{Heide1984271}.
Subsequent high energy electrons incident on the ice will see, and perturb,
the atomic configurations resulting in a Markov-like process for the
evolution of the atomic configuration.  In reality possible atomic configurations 
in amorphous ice are strongly constrained by the ice rules of 
Bernal and Fowler\cite{bernal_theory_1933}.  The simplest
approximation for the transition 
probability of atom positions in going from one conformation to the next 
is to use a Gaussian distribution depending solely on the number of high energy
electrons incident per unit area between the configurations .
In particular if after $n$ electrons per unit area the position of the $i$-th water molecule is 
${\bm x}_{i}(n)$, after a further $d$ electrons per unit area 
the probability distribution for its position has a Gaussian distribution
\begin{equation}
\begin{split}
P(&{\bm x}_{i}(n + d) : {\bm x}_{i}(n) ) = \\ &\frac{1}{ (2\pi d \sigma_{0}^2)^{3/2} }
\exp \bigg\{  -  \frac{ | {\bm x}_{i}(n+d) - {\bm x}_{i}(n) |^2}{2 d  \sigma_{0}^2 } 
	\bigg\}
\end{split}
\label{eqn:AGAUSS}
\end{equation}
in which $\sigma_{0}^2$ is the mean squared displacement 
in a given direction in response to
a single incident electron per unit area.

Assuming an image consists of $M$ frames and has  a total exposure of
$D$ electrons per unit area.
The power spectrum, $S(\bm u )$,
at spatial frequency $\bm u$ in terms of the Fourier components, $f_i({\bm u})$, of the
$i$-th frame is
\begin{equation}
        S({\bm u}) \propto  {<}| \sum_{i=1}^M f_i({\bm u})|^2{>} = \sum_{i,j=1}^M {<}f_i^{*}({\bm u}) f_j({\bm u}){>}
\label{eqn:SumS}
\end{equation}
where ${<}\dots{>}$ denotes an ensemble average.
Since the average number of incident electrons per unit area in each frame 
is $d=D/M$, using Eq.~\eqref{eqn:AGAUSS} the $|i-j|d$ incident electrons between
$i$-th and $j$-th frames will randomly displace the water molecules so that
\begin{equation}
         {<}f_i^{*}({\bm u}) f_j({\bm u}){>} \propto
d^2 F_0 ( { \bm u })^2 \exp( -\alpha_u | i - j | d )
\label{eqn:SumDef}
\end{equation}
in which
$\alpha_u = 2\pi^2 \sigma^2_0 u^2$ and $\sigma^2_0$ is the induced mean squared
motion of a water molecules per incident electron though a  unit area while
$F_0( { \bm u } )^2$ is defined as
\begin{equation}
F_0({\bm u } )^2 \equiv  \frac{1}{Md^2}\sum_{i=1}^M {<} | f_i({\bm u})|^2 {>}.
\label{FU}
\end{equation}
Substituting Eq.~\eqref{eqn:SumDef} into  Eq.~\eqref{eqn:SumS} and summing over $i$ and $j$ gives
\begin{equation}
\begin{split}
       &S({\bm u}) =  d^2F_0({\bm u})^2  \,\,\,\,\times \\
&\bigg(  
  \frac{2\mathrm{e}^{-\alpha_u d}}{ (1 -  \mathrm{e}^{-\alpha_u d })^2}
 \lbrace ( 1 - \mathrm{e}^{-\alpha_u d  }) M
+ \mathrm{e}^{-\alpha_u Md }  - 1
\rbrace + M
\bigg).
\end{split}
\label{eqn:AllSum}
\end{equation}
In cases  where there
is very little induced movement
$\alpha_u \rightarrow 0 $ so that
$S({\bm u}) \rightarrow M^2d^2 F_0({\bm u } )^2 $ and since $d=D/M$ we have
$S({\bm u}) \propto D^2$.
On the other hand if there is no correlation between frames
then $\alpha_u \rightarrow \infty $ and
$S({\bm u}) \rightarrow Md^2 F_0({\bm u } )^2$.

For an exposure with a total of $D$ electrons but where 
the molecules are continuously moving
we can let $M\rightarrow \infty$ and 
the summation in
Eq.~\eqref{eqn:SumS} goes over to an integral giving
\begin{equation}
        S({\bm u}) = 2 F_0({\bm u})^2 [ \alpha_u D + \exp(-\alpha_u D) -1 ]/\alpha_u^2
\label{eqn:IntegralS}
\end{equation}
which in the limits of small and large $\alpha_u$ goes
to $D^2F_0({\bm u})^2$ and $ 2DF_0({\bm u})^2/\alpha_u$, respectively.

\subsection{Estimation of beam induced movement from dose fractionated images}
The amount of movement resulting from an 
incident electron can be estimated from the variation in
Thon ring signal as a function of the number of incident electrons in an image.
When the induced motion is smaller than a given spatial frequency
the Thon ring modulation at that spatial frequency is expected to grow quadratically 
with the number of incident electrons in an exposure but as the induced motion 
becomes comparable with the spatial frequency the amplitude should grow more slowly.   
From Eq.~\eqref{eqn:IntegralS} the 
power spectrum, $S(u)$, at a spatial frequency $u$, 
is expected to vary with total number of electrons per unit
area, $D$,  as
\begin{equation}
	S(u) \propto  2[ \alpha_u D + \exp(-\alpha_u  D) -1 ]/
	\alpha_u^2.
\label{eqn:SU}
\end{equation}
in which $\alpha_u = 2\pi^2\sigma_0^2u^2$, and 
$\sigma_0^2$ is the mean-squared movement in a given direction  
per incident electron passing through a unit area.
The actual Thon ring modulation
does not appear explicitly in this equation but 
the Thon ring modulation enables the signal from the ice to be identified
over the background noise in the measured power spectra.
The measured spatial frequency dependence of $S(u)$ will also include that of
the detector but for the Falcon~II detector this can be removed by using 
the noise whitened power spectra, $W(u)$ as defined in Eq.~\eqref{eqn:WHITE_POWER_SPECTRA}.

From a single dose fractionated 
exposure it is possible to estimate $\sigma_0^2$ through the
dose dependence present in Eq.~\eqref{eqn:SU}.
If an the exposure consists of 
$M$ frames each with an average of $d$ elections per unit area,
the result of summing the original frames in blocks of $m$ 
is to produce $M/m$ frames with on average $md$ electrons per unit area.
Denoting the sum of the resulting 
$M/m$ noise whitened power spectra by
$W_{M,m}(u)$ and 
using  Eq.~\eqref{eqn:SU}  gives
\begin{align}
W_{M,m}(u) \propto  
\frac{ 2 D_{\text{tot}} }{\alpha_u} 
\big(  
		\alpha_{u} md 
 + \exp(- \alpha_{u} md )  - 1 \big)  / \alpha_{u}md
\label{eqn:Wdm}
\end{align}
in which $ D_{\text{tot}} = Md$.
The behaviour of 
Eq.~\eqref{eqn:Wdm} with $m$ depends only
on $z = \alpha_u m d$ and is described by the functional form 
\begin{equation}
g(z) = (z + e^{-z} -1)/z.
\label{eqn:GZ}
\end{equation} 
This is a monotonically increasing function that initially 
varies linearly with $z$ but plateaus towards 1
for large $z$. 
By fitting the non-linear behaviour of
$W_{M,m}(u)$ as a function of  $m$ it is possible to
estimate $\alpha_u$, and hence obtain $\sigma_0^2$.

\section{Experimental}

Quantifoil R1.2/1.3 Cu/Rh grids were used to make a thin film sample 
of double distilled water (18 M$\Omega$cm conductivity). 
The grids were glow-discharged in air for 30-40 seconds and 3 $\mu$l 
of water applied 
in an environmental plunge-freeze apparatus \cite{JEMT:JEMT1060100111}.  Grids 
were blotted for 4-6 seconds and rapidly frozen in liquid ethane. 
The grids were transferred to Krios cartridges and imaged using a 
Falcon~II direct electron detector on a FEI Titan Krios
operated at 300 keV. A nominal magnification of 
75,000x corresponding a calibrated value of
134,600x was used which results in a 1.04 \AA\/ sampling with the 
14 $\mu$m pixels of the Falcon II detector.
Imaging was performed using nano-probe mode
with parallel illumination using a 70 $\mu$m C2 aperture and no
objective aperture. A beam slightly larger and centred on the Quantifoil hole was used.
Astigmatism and beam tilt 
correction were performed at the imaging magnification. 
Images were recorded at 18 frames/sec for either 8 or 17 seconds with
either 2.33 e$^{-}$/frame, or 0.85 e$^{-}$/frame, respectively.  
The electron exposure per frame was determined from the screen current and 
average signal from the Falcon detector, with these being calibrated against 
a picoammeter as described previously \cite{mcmullan_comparison_2014}.

The amorphous carbon control specimen was made in an
Edwards 301 vacuum coating unit by 
evaporating carbon from an arc onto mica. 
The thickness of the carbon film was estimated to be $150\,$\AA\/ from
the optical density of the carbon deposited on adjacent piece of filter paper. 
The carbon film was floated off the mica using a water bath 
and placed onto a Cu/Rh grid.  The continuous carbon grid
was examined using the Atlas component of the FEI EPU software.
From the Atlas an area of the carbon film was selected 
in which the carbon film had no holes or wrinkles both 
in the grid square and surround grid squares.
After pre-irradiation of the carbon, images were recorded for 1.5 seconds
with the same magnification and number of incident electrons per frames 
as used for the ice  sample.
The stage was moved to adjacent 
areas in a raster (using $0.4\,\mu$m steps in X followed by $0.4\,\mu$m steps in Y). 
A total of 25 images from different but adjacent areas of carbon film were collected with 
the defocus for all images set as close as possible to a nominal ~5000 \AA\/ by manually
adjusting the focus at each position. Later 
evaluation using CTFFIND3 \cite{Mindell2003334} showed that the defocus 
was actually $5990\,$\AA, with $200\,$\AA\/ of astigmatism and ${\pm}90\,$\AA\/ defocus variation 
over the 25 images.  

Images were recorded using the full $4\mathrm{k}{\times} 4\mathrm{k}$ output
of the Falcon~II detector with the individual frames of all the images being
captured using an in-house data capture setup. In processing the exposures the
individual frames were not aligned computationally.

\section{Results}


\begin{figure}[t]
\centerline{\includegraphics[width=0.5\textwidth]{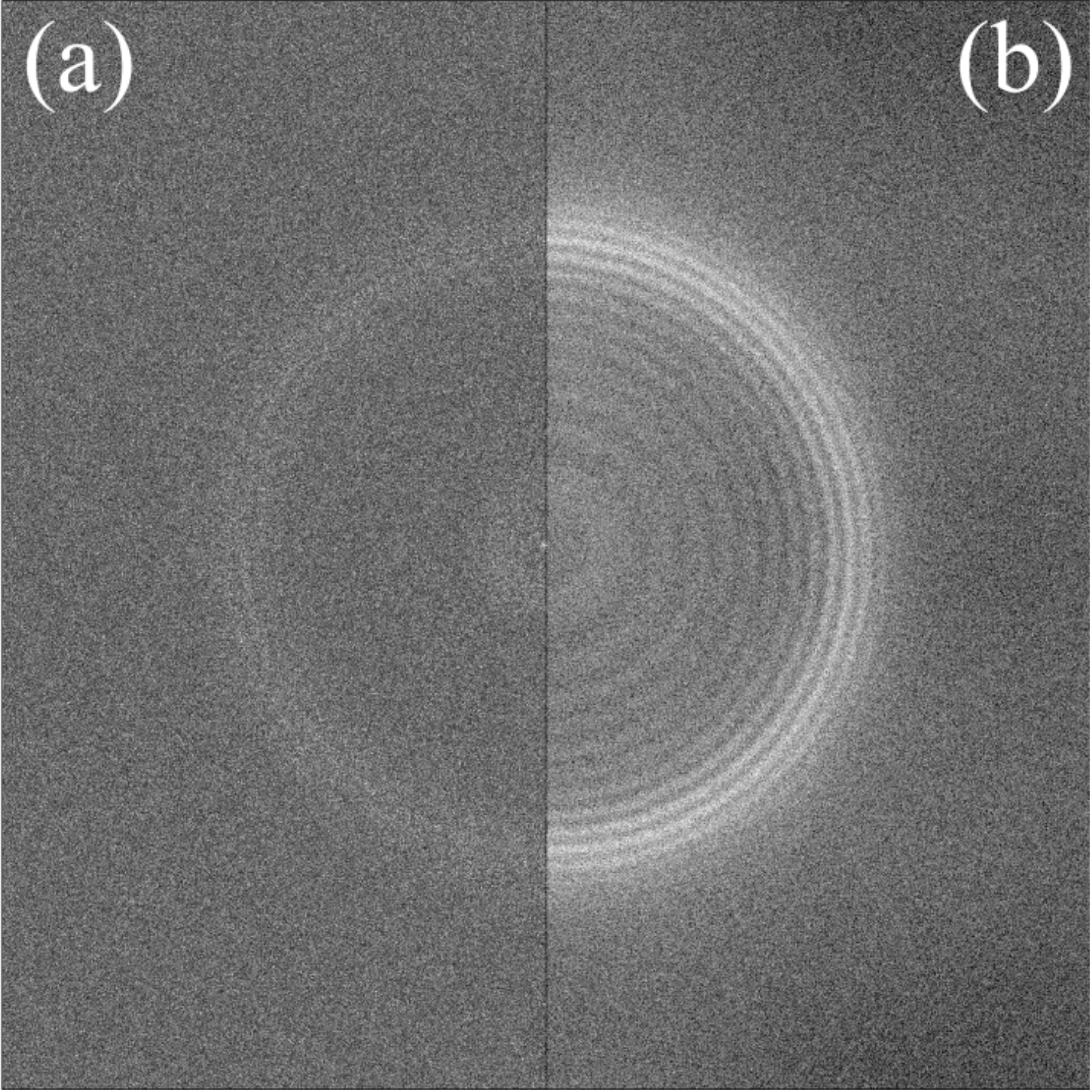}}
\caption{Thon rings from a dose fractionated exposure (image \#230804) of amorphous ice obtained
using either (a) the NWPS of the sum of the frames,  or (b) the sum of
the NWPS from each frame. 
The exposure consisted 141 frames at a dose of $2.33\,$e$^{-}$/pixel using 300 keV electrons,
a defocus of $7700\,$\AA\/ and a pixel sampling of $1.04\,$\AA\/. The edge of the transform is at
$1/2.08\,$\AA$^{-1}$ and the strong ring pattern in (b) is centred around 1/3.7 \AA$^{-1}$.
In both (a) and (b) the the images were first scaled so that the RMS noise level was 1 and 
then the lower and upper grey limits set at -0.7 and 1.3 about the mean values.
\label{fig:ICE_THON}}
\end{figure}

Fig.~\ref{fig:ICE_THON} shows two types of noise whitened power spectra, $W({\bm u})$,
\cite{mcmullan_comparison_2014}
from a dose fractionated exposure of pure amorphous ice consisting
of 141 frames with on average 2.33 e$^{-}$/pixel per frame.
The spectrum from the sum of all the 141 frames is shown in 
Fig.~\ref{fig:ICE_THON}(a) and with such a relatively high 
dose (300 e$^{-}/$\AA$^2$) it is possible to see faint Thon rings.
Fig.~\ref{fig:ICE_THON}(b) 
shows the sum of the 141 noise whitened power spectra of the individual frames. 
Exactly the same images were used but in the sum of the power spectra 
of the individual  images  the Thon rings that are barely visible in 
Fig.~\ref{fig:ICE_THON}(a) are clearly visible and can be seen to extend out beyond
$3.4\,$\AA\/ resolution.
The $\sqrt{141}$ reduction in the noise
from averaging the 141 individual spectra enables the strength of the Thon rings 
in individual frames to be seen clearly.
The strength of this signal indicates that the Thon rings arise 
from the intrinsic bulk water and not
from impurities absorbed from the adjacent carbon film or contamination occurring 
during the blotting and transfer steps.

The behaviour of the circularly averaged power spectra
and circularly averaged noise whitened power spectra are compared
in Sec.~\ref{sup:compare} of the  supplementary material.
The flat, featureless backgrounds in 
Fig.~\ref{fig:ICE_THON}(a) and 
Fig.~\ref{fig:ICE_THON}(b) illustrate the success of 
the noise whitening procedure.
The measured average value and noise in the noise whitened power spectra 
of Fig.~\ref{fig:ICE_THON} are  
within 5\% of the values predicted by
Eqs .~\eqref{eqn:WM} and 
\eqref{eqn:NOISE} using $n=2.33$ and DQE(0)= 0.5.
The probability distribution based on all the values in the 
141 noise whitened power spectra 
used in Fig.~\ref{fig:ICE_THON}(b) is given in Sec.~\ref{sup:hist} of the
supplementary material. The distribution is well described by
Eq.~\eqref{eqn:PROB} with a measured value of $\Gamma = 4.91$ that is 
within 5\% of an estimated based on Eq.~\eqref{eqn:WX} again using
$n=2.33$ and DQE(0)= 0.5.


To illustrate the origins of the difference between 
Fig.~\ref{fig:ICE_THON}(a) and Fig.~\ref{fig:ICE_THON}(b) we
carried out two control experiments using amorphous carbon film.  
The amorphous carbon film was first pre-irradiated as 
even atoms in films of carbon prepared by evaporation 
from a carbon arc {\sl in vacuo} move when initially irradiated. 
The amount of movement decreases with exposure but 
$\sim$100 e$^{-}$/\AA$^2$ is sufficient 
to effectively stabilise a film,  i.e., the observed Thon ring pattern 
showed no drift and was stable.
As 300 keV electrons can cause displacement 
damage\cite{egerton_radiation_2004}, there was some
residual movement but 
relative to the initial movement this was negligible.


\begin{figure}[t]
\centerline{\includegraphics[width=0.5\textwidth]{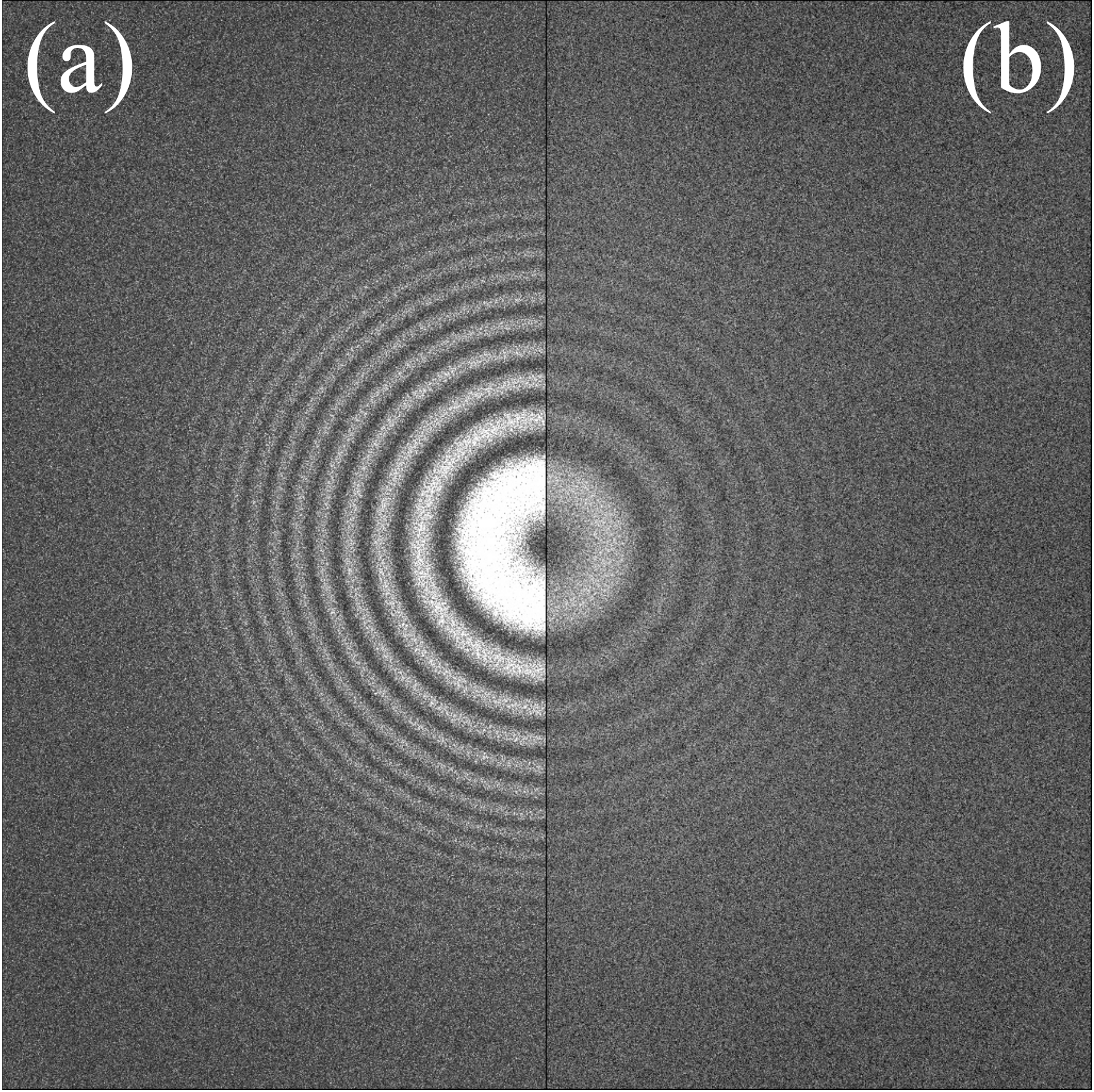}}
\caption{
Comparison from a dose fractionated exposure of an area of pre-irradiated carbon
of (a) the NWPS obtained from the sum of all the frames and (b) the
sum of the NWPS of individual frames. 
The exposure consisted of 25 frames with on average
2.33 e$^{-}$/pixel per frame recorded using 300 keV electrons, a 5990~\AA\/ defocus and 
a 1.04~\AA\/ pixel sampling.
The grey scales of the images were set as in Fig.~\ref{fig:ICE_THON}.
\label{fig:CARBON_THON}}
\end{figure}

In the first control experiment, the initial 25 frames from a 1.5 second exposure
of an area of pre-irradiated carbon were used. 
The magnification and number of incident electrons per frame were 
the same as in Fig.~\ref{fig:ICE_THON}.  The corresponding 
noise whitened power spectra from the summed image and the sum of the individual power
spectra are given in Fig.~\ref{fig:CARBON_THON}(a) and 
Fig.~\ref{fig:CARBON_THON}(b), respectively.
In  contrast to Fig.~\ref{fig:ICE_THON}, the Thon rings are now stronger in the power spectrum
of the sum of the frames with the relative strengths being essentially 
what is expected from
25 images of the same object.


\begin{figure}[t]
\centerline{\includegraphics[width=0.5\textwidth]{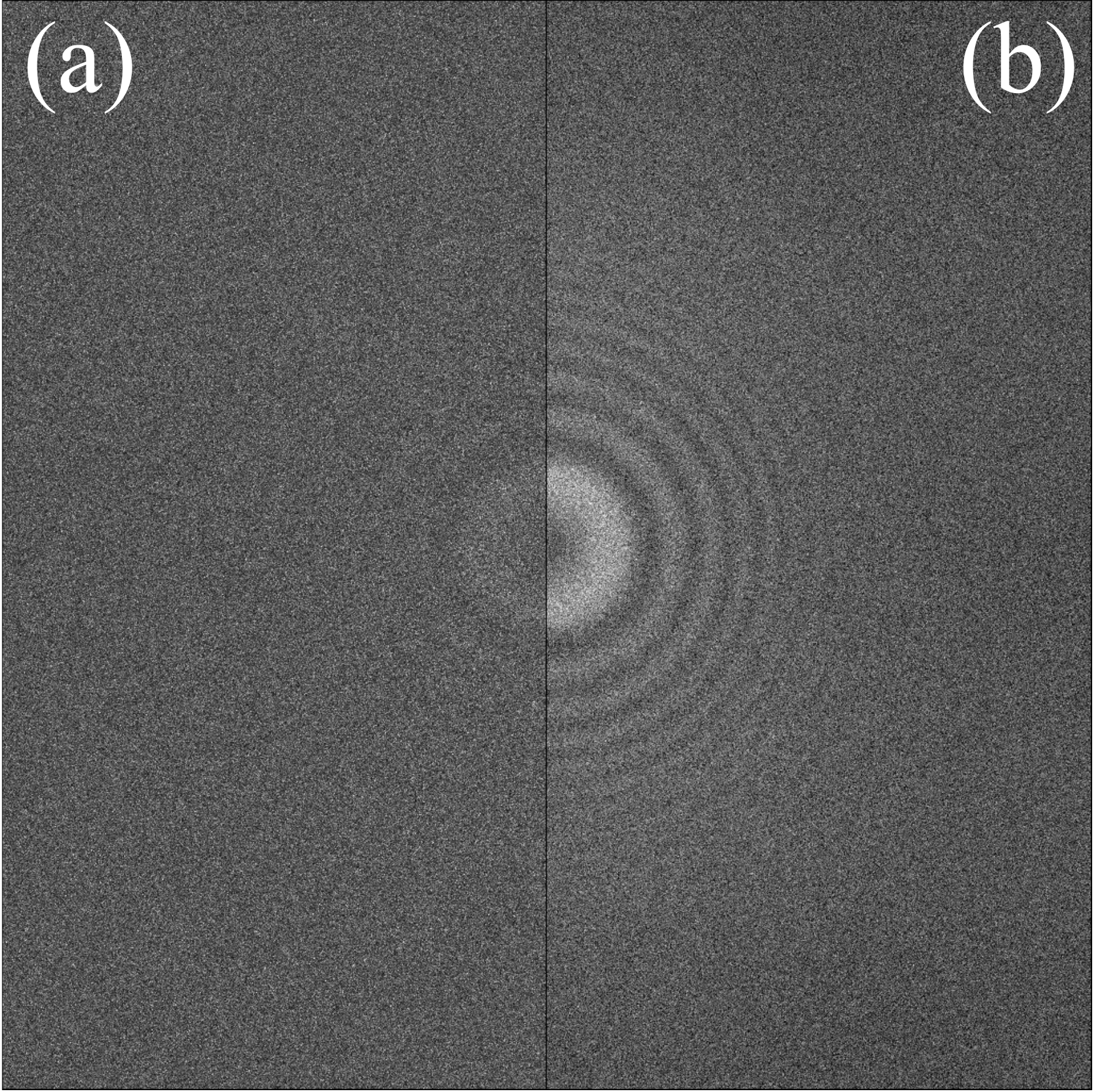}}
\caption{
Comparison from a composite exposure 
of (a) the NWPS obtained from the sum of all the frames and (b) the
sum of the NWPS of the individual frames. 
The composite image consisted of 25 single frames taken from different 
dose fractionated images of non-overlapping areas in a region of pre-irradiated amorphous carbon.  
The imaging conditions were the same as in  Fig.~\ref{fig:CARBON_THON} with 
the grey scales of the images set as in Fig.~\ref{fig:ICE_THON}.
\label{fig:CARBON_RANDOM_THON}}
\end{figure}

In the second control experiment, a 
series of 25 images like that in Fig.~\ref{fig:CARBON_THON} were taken
at adjacent but non-overlapping areas of the pre-irradiated carbon.
As there was a slight variation in 
the height of the carbon film 
at the different locations the objective lens current
was adjusted in order to keep
the variation in the defocus of the images to within ${\pm}90\,$\AA. 
A composite image of 25 frames was then generated 
by taking one frame from each of the 25 images.
As in Figs. \ref{fig:ICE_THON} and \ref{fig:CARBON_THON} the 
corresponding power spectra from
the Fourier transform of the sum of the frames from the composite image is 
given in 
Fig.~\ref{fig:CARBON_RANDOM_THON}(a) while that of the sum of the
power spectra of the individual frames in the composite image is shown in 
Fig.~\ref{fig:CARBON_RANDOM_THON}(b).
Like Fig.~\ref{fig:ICE_THON}, the
Thon rings in Fig.~\ref{fig:CARBON_RANDOM_THON} are more visible in the sum of 
the power spectra than in the power spectrum
of the summed image. As expected Fig.~\ref{fig:CARBON_THON}(b) and 
Fig.~\ref{fig:CARBON_RANDOM_THON}(b) are almost identical since they 
both consist of the sum of 25 power spectra from images 
of areas with the same thickness of stabilised carbon taken with essentially the 
same defocus and number of electrons.

In Fig.~\ref{fig:CARBON_THON}, the carbon atom positions are
effectively the same in all the frames while in
Fig.~\ref{fig:CARBON_RANDOM_THON} there no  correlations between
atom positions in the different frames. 
The similarity of Fig.~\ref{fig:ICE_THON} to Fig.~\ref{fig:CARBON_RANDOM_THON}
indicates that water molecules in amorphous ice are moving to uncorrelated 
positions during an exposure.

\begin{figure}[t]
\centerline{\includegraphics[width=0.5\textwidth]{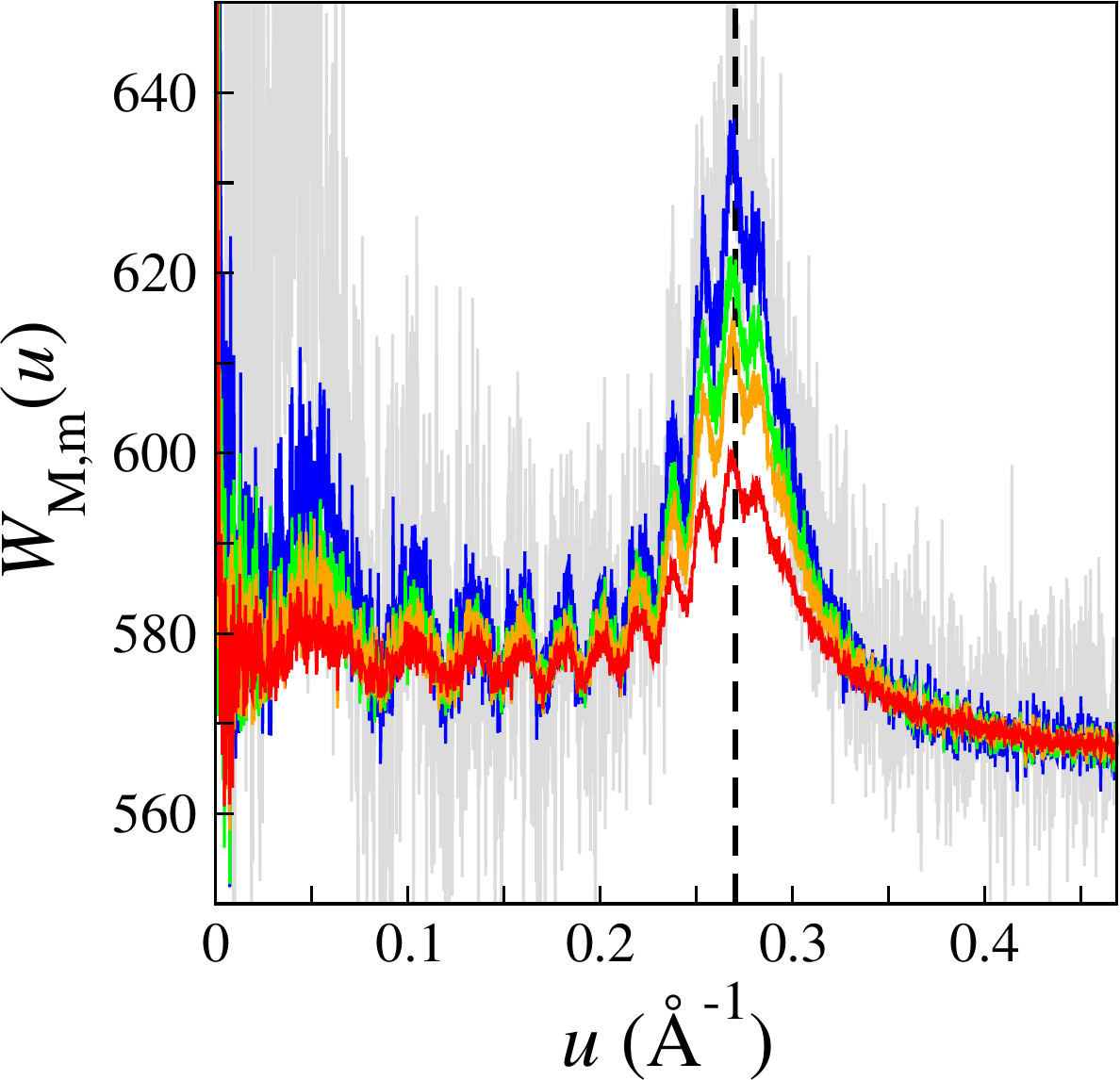}}
\caption{Circularly averaged power spectra, $W_{120,m}(u)$ as a function of spatial frequency, $u$,
from  120 frames of image \#230804 used in Fig.~\ref{fig:ICE_THON}
showing the behaviour with $m$. 
Results for $m=1$ (red), $m=2$ (orange), $m=3$ (green),
$m=8$ (blue), $m=120$ (grey) are shown. 
Exposure \#230804 had 2.33 e$^{-}$/pixel/frame and 
using Eq.~\eqref{eqn:WM} with a DQE of 0.5 for the Falcon~II gives an expected average background $\Gamma =  550$.
The expected noise in the limit of zero spatial frequency from Eq.~\eqref{eqn:WM} 
is $50\sqrt{m}$ but this decreases with the square root of the spatial frequency 
as more values are included in the circular average. The vertical dashed line indicates the position of 
1/3.7\AA\/ data used in Fig.~\ref{fig:FIT}.
\label{fig:PWR} }
\end{figure}

Fig.~\ref{fig:PWR} shows the circularly averaged values
of $W_{M,m}(u)$ at selected values of $m$ obtained from the 
dose fractionated image series 
of amorphous ice in exposure \#230804 used in Fig.~\ref{fig:ICE_THON}. 
For simplicity in factorisation, only the first 120 of 141 frames were used. 
The average background value, the magnitude of the associated noise and their variation
with $m$ are in agreement with the predictions of Eq.~\eqref{eqn:WM} and 
Eq.~\eqref{eqn:NOISE}.  There are some residual correlations within, and between, frames
that lead to small systematic variations in $W_{M,m}(u)$ with $m$.  These were
removed by adding a small  $m$ dependent shift for each $m>1$ to ensure
that at the Nyquist frequency (in this case 1/2.08\AA) 
$W_{M,m}(u) = W_{M,1}(u)$.
The maximum shift required was for $m=120$ and corresponded to 5\% of the 
background.

\begin{figure}[t]
\centerline{\includegraphics[width=0.5\textwidth]{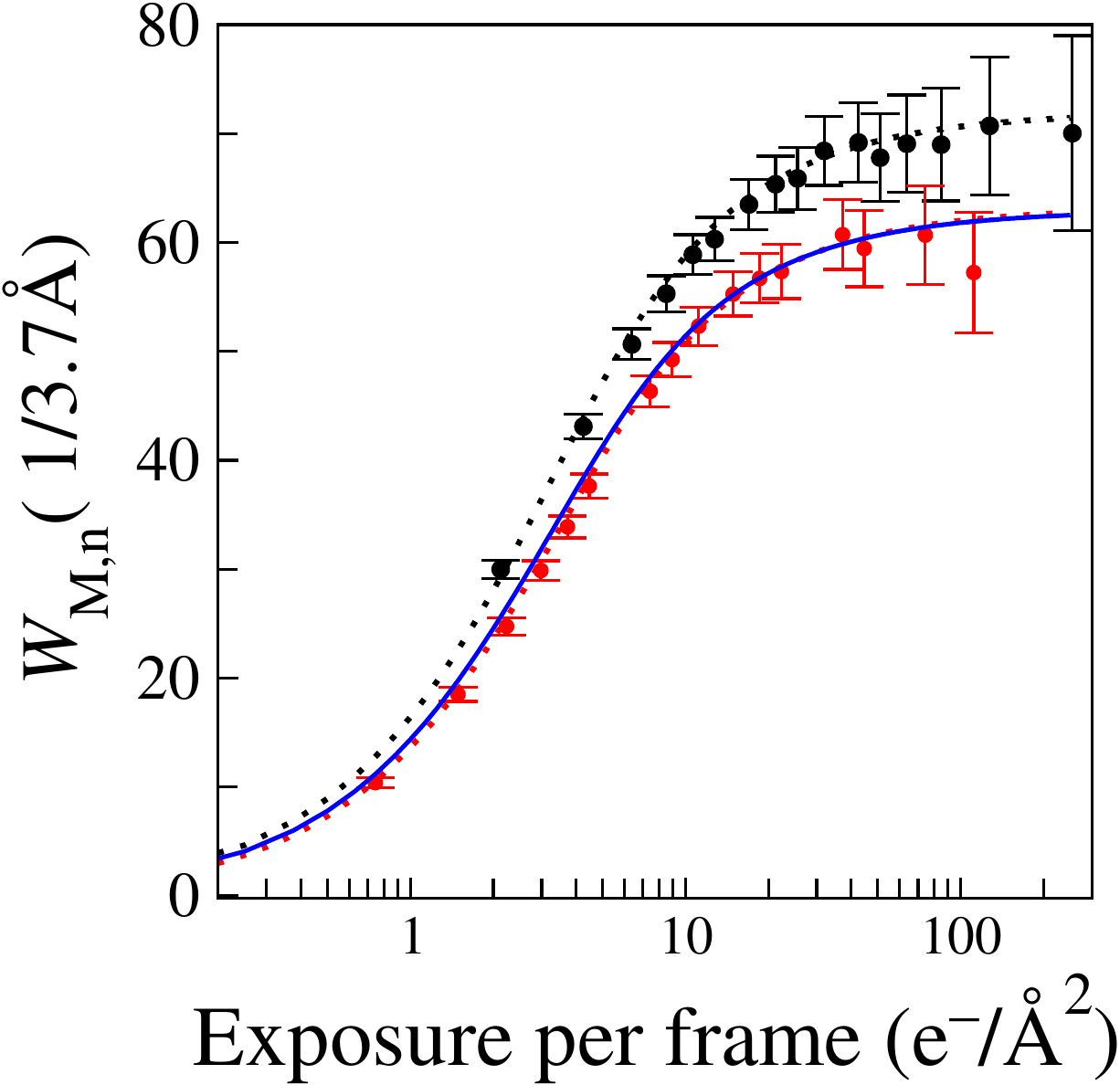}}
\caption{Measured Thon ring modulation from 
the circularly averaged $W_{M,m}({\bm u})$  at 
$u=1/3.7$\AA\/ as a function of the number of electrons per unit area in the summed frames. 
The results shown using black bullet ($\bullet$)  are
from the first 120 frames of exposure 
\#230804 with 2.33 e$^{-}$/pixel/frame.
The results shown using red bullet ($\color{red}\bullet$) are 
from the first 300 frames of exposure \#171528 taken
with the same magnification and defoucs but over
15 seconds using 0.85 e$^{-}$/pixel/frame 
on a different amporphous ice sample.
The red and black dotted lines show fits
to the corresponding experimental data using 
Eq.~\eqref{eqn:Wdm}.
The solid blue curve
lies almost on top of the dotted red curve and 
is simply 0.87 times the dotted black curve.
The error bars are
based on circularly averaged estimates using
Eq.~\eqref{eqn:NOISE}. 
\label{fig:FIT}}
\end{figure}

Fig.~\ref{fig:FIT} shows the circularly averaged values 
of  $W_{M,m}({\bm u})$ as a function of $m$
at $u=1/3.7$\AA\/ 
(indicated by the vertical dashed line in Fig.~\ref{fig:PWR})
from 120 frames of exposure \#23804.
Also shown in Fig.~\ref{fig:FIT} are results 
from a  dose fractionated image (\#171524) consisting  of 300 frames 
from of a different amorphous ice sample using 
the same magnification as image \#23804 but with
an exposure rate of 0.85 e$^{-}$/pixel/frame.
The dotted lines in Fig.~\ref{fig:FIT} are fits
to the measured results using Eq.~\eqref{eqn:Wdm}  and 
two 
adjustable parameters: $\sigma_0^2$ and a scale factor.
The scale factor includes the electron optical terms and is 
proportional to the total number of electrons in the exposure, 
the scattering strength 
and thickness of the amorphous ice sample.  Fits to 
successive dose fractioned images of a fixed area of amorphous ice 
give  essentially the same value for $\sigma_0^2$ 
but the scale factor decreases as the 
sample is thinned by radiolysis.  
The fitted  values of $\sigma_0^2$ shown in 
Fig.~\ref{fig:FIT}  
for images \#230804 and \#171524 are
0.38 and 0.35 \AA$^2$/(e$^{-}/$\AA$^2$), respectively.
There are many sources of systematic error in this analysis and 
at best these values for $\sigma_0^2$ should be considered as estimates.
Based on these estimates the radiation damage resulting from one 
incident 300 keV e$^{-}$/\AA$^2$ is expected to induce 
a total mean squared motion of
$\sim 1.1$  \AA$^2$ in 
the water molecules of the sample.


For a dose fractionated exposure of $M$ frames the
Thon ring signal 
given  by Eq.~\eqref{eqn:GZ}  (and illustrated in Fig.~\ref{fig:FIT})
initially increases linearly with $m$  but plateaus at higher $m$. 
As the corresponding noise, given by  Eq.~\eqref{eqn:NOISE}, 
grows as $\sqrt{m}$ there is an optimal $m$, 
and hence dose per image, with which
to observe Thon rings.
Dividing Eq.~\eqref{eqn:GZ} by $\sqrt{z}$ and setting the derivative 
with respect to $z$ to zero gives a transcendental equation with 
zero at 2.149. For a given resolution, $u$,  the
optimal number of electrons per unit area, $d_\text{opt}$ in a 
frame for observing Thon rings is 
\begin{equation}
	d_\text{opt} = 2.149/2\pi^2\sigma_0^2 u^2.
\label{eqn:PEAK}
\end{equation}
The peak is not very sharp but 
for a given dose the Thon rings around 1/3.7\AA\/ from amorphous ice (assuming 
$\sigma_0^2 = 0.37\,$\AA$^2$/(e$^{-}$/\AA$^2$)) will
have the highest signal to noise ratio
if each frame has, or successive frames are
grouped into blocks with an average exposure of 
$\sim 4.0\,\mathrm{e}^{-}$/\AA$^2$.



\begin{figure}[th]
\centerline{\includegraphics[width=0.5\textwidth]{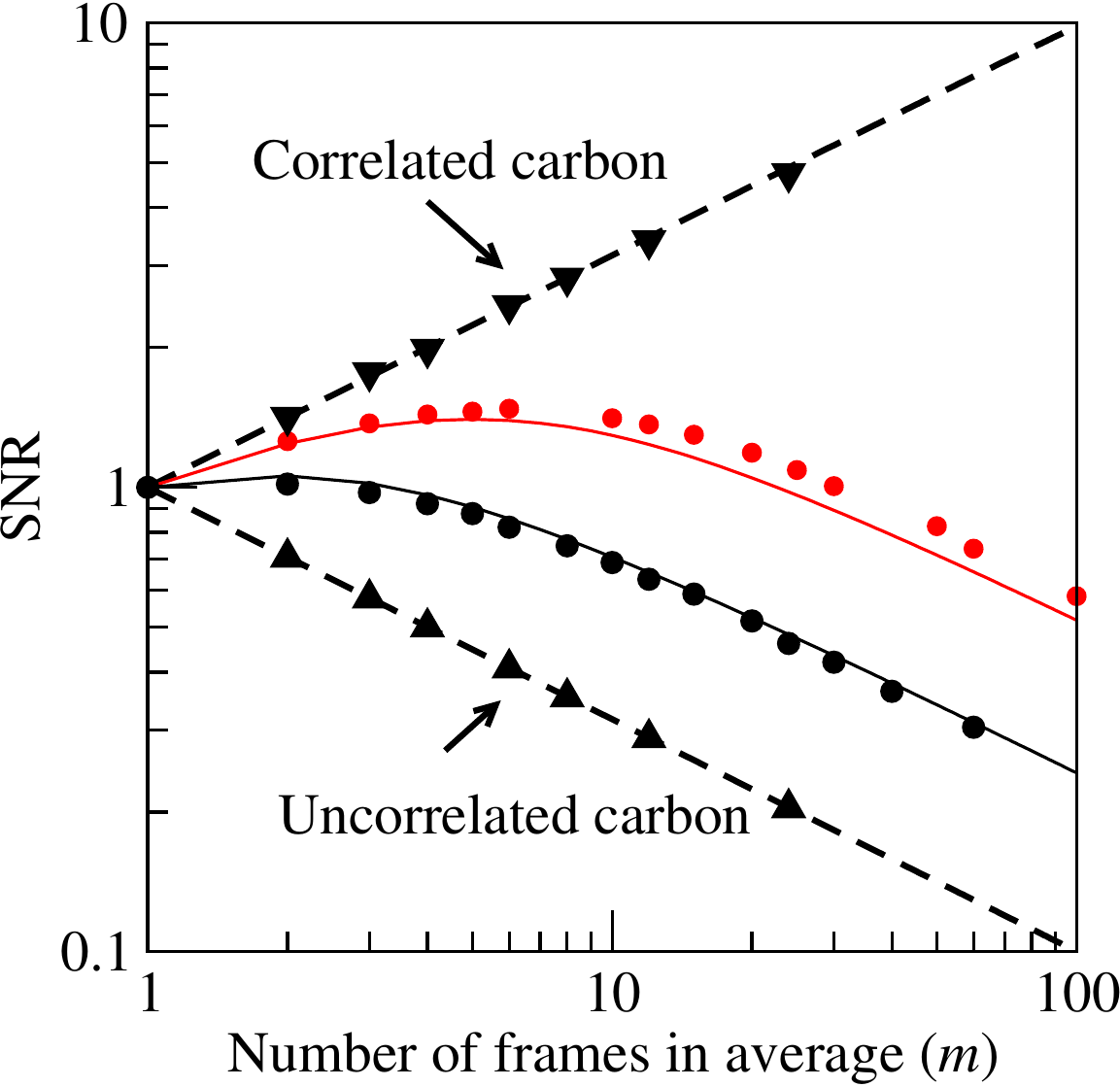}}
\caption{Relative signal-to-noise ratio of Thon rings in the sum of noise whitened
power spectra, $W_{M,m}/W_{M,1}$, from a dose 
fractionated exposure of $M$ frames as a function of the 
number of successive images in each power spectrum, $m$. 
Results at $u=1/9.0\ $\AA\/ from 24 frames of the pre-irradiated carbon
images used for Fig.~\ref{fig:CARBON_THON}  ($\blacktriangledown$)
and Fig.~\ref{fig:CARBON_RANDOM_THON} ($\blacktriangle$) are shown. Results 
at $u=1/3.7\ $\AA\/ from 120 frames of
image \#23804 ($\bullet$) 
with 2.33 e$^{-}$/pixel and 
300 frames of image \#171524 
($\color{red}\bullet$)
with 0.85 e$^{-}$/pixel 
of Fig.~\ref{fig:FIT} are shown.  
The results using Eq.~\eqref{eqn:NOISE} for the noise and 
Eq.~\eqref{eqn:GZ}) for the signal
with $\sigma^2_0= 0.37 $\AA$^4$/e$^{-}$ are shown as solid lines.
\label{fig:RH}}
\end{figure}

Finally, while Fig.~\ref{fig:FIT} gives the amplitude of the observed Thon ring
modulation in  $W_{M,m}(u)$ as $m$ (and hence the exposure per image) is varied,
it is also useful to look at how the corresponding visibility of the Thon ring modulation, i.e., the 
signal-to-noise ratio, varies.
Fig.~\ref{fig:RH} shows a log-log plot of the variation in 
the signal-to-noise relative to the
corresponding case with $m=1$  in the noise whitened power spectra as a function of $m$.
Results are shown for both the amorphous ice exposures shown in Fig.~\ref{fig:FIT} as well as the
pre-irradiation carbon cases shown in Fig.~\ref{fig:CARBON_THON} and \ref{fig:CARBON_RANDOM_THON}.
The results for amorphous ice fall between two extremes, given by 
a line varying as $\sqrt{m}$ for the case where the images are of the same object in each frame 
(such as in Fig.~\ref{fig:CARBON_THON} for the images of the same area of 
pre-irradiated carbon), and in the other extreme of 
a line varying as $1/\sqrt{m}$ for the case where the images are of uncorrelated objects 
in the different frames (such as in Fig.~\ref{fig:CARBON_RANDOM_THON} made up of images of different
areas of amorphous carbon).
The $\sqrt{m}$ and $1/\sqrt{m}$ limits indicated  by the dotted lines in 
Fig.~\ref{fig:RH} simply follow from noting
$W_{M,m}$ as defined in \eqref{eqn:Wdm},
varies as  $m$ in the limit where there is no movement between frames but goes to a constant 
when there is no correlation between frames while the 
noise given by  Eq.~\eqref{eqn:NOISE} varies as $\sqrt{m}$. 
The signal-to-noise ratio curves for both exposure \#23804 and \#151524 
initially vary as $\sqrt{m}$ but make the transition to 
$1/\sqrt{m}$ for large $m$.  As expected 
from Eq.~\eqref{eqn:PEAK} in going from behaving as $\sqrt{m}$ to $1/\sqrt{m}$ 
the amorphous ice curves reach a peak at $m$ corresponding to 
$\sim 4 \,$e$^{-}$/\AA$^{2}$.
The solid lines in
Fig.~\ref{fig:RH} show the predicted behaviour of $g(z)/\sqrt{z}$ in which $g(z)$ is
defined in Eq.~\eqref{eqn:GZ} with $z = 2\pi^2 \sigma_0^2 u^2 m d $ at
$u=1/3.7$\AA\/  with $\sigma_0^2= 0.37$\AA$^2$/(e$^-$/\AA$^2$) and 
$d = 2.33$ e$^{-}$/frame, (black {\color{black}---}), or 
$d = 0.85$ e$^{-}$/frame (red {\color{red}---}).

\section{Discussion}

The recent introduction of higher performance electron detectors has allowed
structures of biological macromolecules to be obtained by single particle
cryoEM to higher resolution, with fewer particles and more easily than
before \cite{Kuhlbrandt28032014}. However, these reconstructions still require
averaging substantially more particle images than predicted to be necessary by
theory \cite{henderson_potential_1995}.
Charging of the specimen and its surroundings as well as physical movement
of the specimen during exposure to the first few e$^{-}$/\AA$^2$ are two 
reasons for this discrepancy.

To observe well defined Thon rings from images 
of amorphous ice the sample must be neither too thick nor too thin.  
If the ice film is too thin, the amplitude of the 
Thon ring signal will simply be too small to be seen amongst  
the background noise. If the ice film is too thick 
the Thon signal will also become weaker, especially at high resolution.  
Part of this reduction comes from the variation in defocus 
as images of molecules at different heights have
Thon ring zeroes at different spatial resolutions, 
and part is due to increased multiple elastic and inelastic 
scattering of the electron beam. 
If an ice film is too thick it is possible to 
use radiolysis to thin the film to the desired thickness by recording a 
succession of images.  
With 300 keV electrons the thickness of an 
ice film is typically reduced by $100\,$\AA\/ for every 
$170\,$e$^{-}$/\AA$^2$ of electron dose  \cite{Wright2006241}.  
From the number of successive images required to completely burn through the 
ice film used in Fig.~\ref{fig:ICE_THON} we estimate
that the image used in Fig.~\ref{fig:ICE_THON} was taken 
when the ice thickness was between 1000 and 1500 \AA. 

The high number of electrons used in 
Fig. ~\ref{fig:ICE_THON} (304 e$^{-}$/\AA$^2$) 
ensures the Thon rings are clearly visible but is not necessary.
With the correct conditions Thon rings can easily seen 
with 26 e$^{-}$/\AA$^2$.  In fact the Thon ring visibility in 
the sum of the power spectra from 12 frames of amorphous ice 
exposure from Fig.~\ref{fig:ICE_THON} 
is comparable to that seen 
in the sum of the power spectra of same number of frames from the 
amorphous carbon results of Fig.~\ref{fig:CARBON_RANDOM_THON} 
(see supplementary Sec.~\ref{sup:carbon_versus_ice}).
The pattern of the Thon rings from amorphous ice differs 
from that of carbon, with
the amorphous ice signal,  as in Fig.~\ref{fig:ICE_THON}, 
strongest around 1/3.7\AA\/ while the corresponding pattern from amorphous
carbon is strongest at low spatial frequencies. The 
$3.7\,$\AA\/ resolution signal in amorphous ice comes from 
planes of next nearest neighbour oxygen atoms of tetrahedrally 
coordinated water molecules and the strength of this signal
indicates that the ice rules of Bernal and Fowler \cite{bernal_theory_1933} 
are applicable in amorphous ice.


The major origin of beam induced  motion in an amorphous ice film  
is from radiolysis rather than knock-on damage \cite{Heide1985151}. 
The rate of mass loss from a thin films is observed to be almost 
independent of film thickness and so most of the radiolysis fragments remain
within the film \cite{Heide1985151}. This so called cage-effect, 
allows many of the radiolysis fragments in ice to recombine but some
such as O$_2$ and H$_2$O$_2$, will accumulate and re-orient the surrounding 
water molecules.
The successful fits shown in Fig.~\ref{fig:FIT} do however support 
the simple Gaussian Markov Process description of Sec.\ref{sec:GAUSS}.
In practice irradiation of a plunge frozen amorphous ice film will also
result in collective motion of local regions due to stress relief within the 
film and charging from the emission of secondary electrons.


From our experimental 
observation we deduce that water molecules, each with a molecular mass of 
$18\,$Da, move an RMS distance of $\sim$1\AA\/ for each e$^{-}$/\AA$^2$.
A typical cryoEM exposure having  $25\, $e$^{-}$/\AA$^2$ will therefore result 
in a RMS displacement of water molecules by ${\sim}5\,$\AA.
In the same way as the thermal motion of water molecules results
in Brownian motion, the beam-induced motion of water molecules will 
also be expected to displace embedded protein molecules.
In a $25\,$e$^{-}$/\AA$^2$ exposure a protein molecule of $100\,$kDa, such as hexokinase, 
embedded in this film of randomly diffusing molecules would be expected to 
have an RMS displacement of ${\sim}1.0\,$\AA\/
(hexokinase with a MW of $100\,$kDa has a 20-40x smaller diffusion 
coefficient than water, and so should move ${\sim}$4-6x less, according to 
the Stokes-Einstein equation for diffusion \cite{Einstein1905} in which 
the diffusion coefficient varies as $1/\mathrm{MW}^{1/3}$). 
For a protein molecule 
of 25 kDa, the beam-induced random motion would be higher at ${\sim}1.5$\AA, 
and for a ribosome of $2.5\,$MDa somewhat lower at ${\sim}0.7$\AA.  
This additional B-factor, which for hexokinase would 
correspond to ${\sim}25\,$\AA$^2$, produces some extra image blurring on top 
of that resulting from other effects such as intrinsic radiation 
damage to the macromolecular assembly.
We can thus conclude that random, 
Brownian type of beam-induced motion of biological structures is 
unlikely to be one of the limiting factors in attaining 
high-resolution structures using single particle cryoEM approaches.  
Only for very small particles 
at resolutions of 2\AA\/ or beyond is this type of beam-induced motion likely 
to be a fundamental limitation.

It can be argued that the observed degree of beam-induced motion 
of water molecules in amorphous ice during an exposure is actually of positive benefit in
cryoEM. If water molecules moved much less during irradiation they would
contribute a strong background, just as there is from carbon films,  that would
decrease the accuracy of the orientation determination in single particle cryoEM. If 
water molecules moved more then so too would the embedded macromolecules 
and the images of these would be more blurred.

\section*{Acknowledgments and declaration of interest}
The authors acknowledge funding from the Medical Research 
Council, grant number U105184322. We would also like to thank
Garib Murshudov and Nigel Unwin for their helpful comments.

\bibliographystyle{elsarticle-num}
\bibliography{mcmullan_thon_ice}

\clearpage





\newcommand{\beginsupplement}{%
  \renewcommand{\thetable}{S\arabic{table}}%
  \setcounter{figure}{0}%
  \renewcommand{\thefigure}{S\Roman{figure}}%
  \renewcommand{\theequation}{S\Roman{equation}}%
  \renewcommand{\thesection}{S\Roman{section}}%
  \renewcommand{\thepage}{S.\Roman{page}}%
  
}
\beginsupplement

\onecolumn
\begin{center}
\textbf{\LARGE{Supplementary material: Thon rings from amorphous ice and implications of beam-induced
Brownian motion in single particle electron cryo-microscopy}}
\end{center}

\begin{figure}[ht]
\begin{minipage}[b]{0.45\linewidth}
\centering
\includegraphics[width=\textwidth]{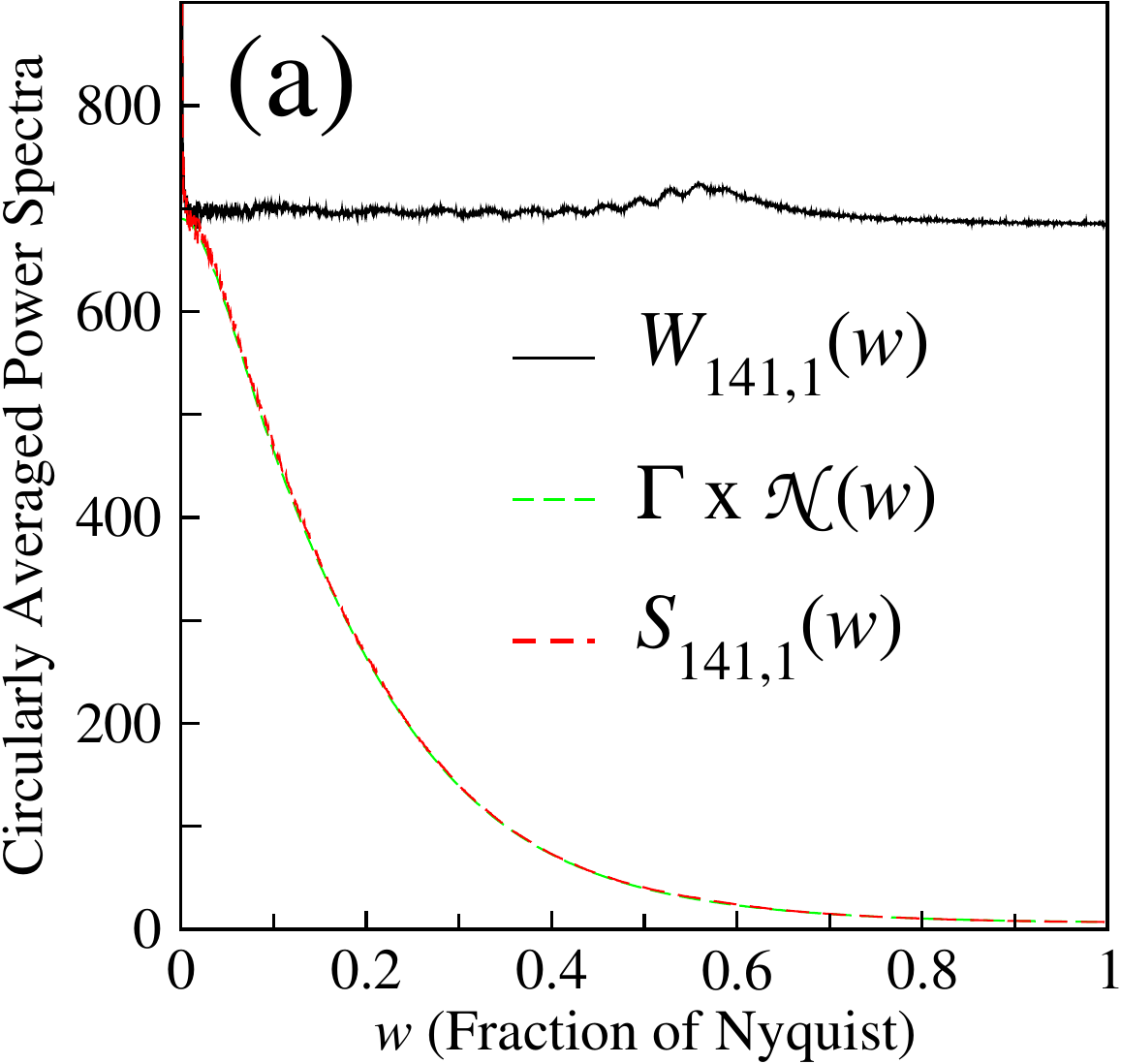}
\end{minipage}
\qquad
\begin{minipage}[b]{0.45\linewidth}
\centering
\includegraphics[width=\textwidth]{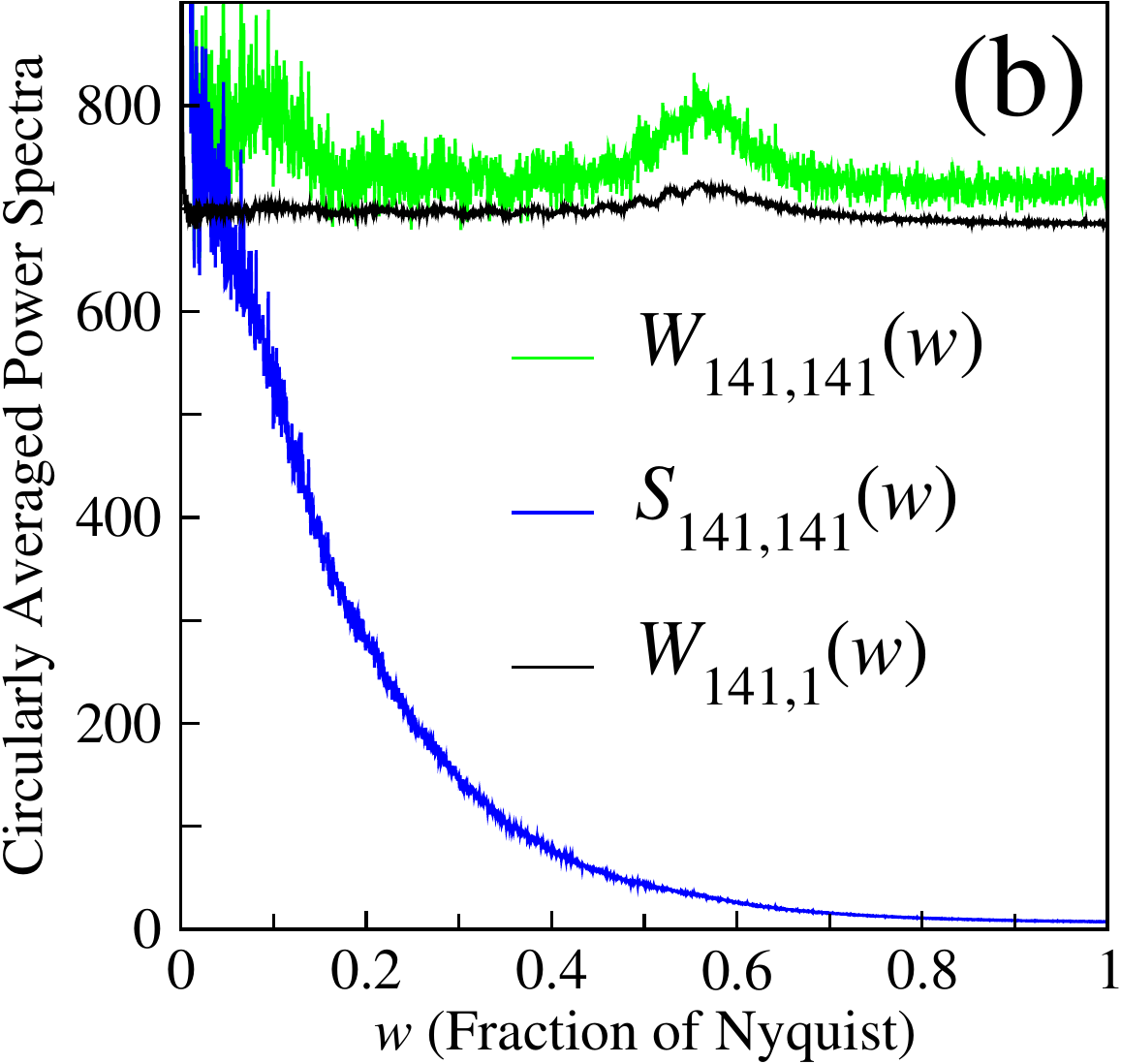}
\end{minipage}
\caption{Comparison of the spatial frequency dependence of the 
circularly averaged sum of power spectra and noise whitened power spectra from the 
141 frame dose fractionated exposure (image \#230804) of amorphous ice used for Fig.~\ref{fig:ICE_THON}.
The spatial frequency, $w$ is measured as a fraction of the Nyquist frequency (1/2.08\AA).
In (a), the sum of the power spectra of the individual images ($S_{141,1}(w)$ and $W_{141,1}(w)$) 
along with $\Gamma {\mathcal N}(w)$ using $\Gamma= 684$ are given.  In (b), the circularly
averaged power spectrum of the sum of the 141 frames 
($S_{141,141}(w)$ and $W_{141,141}(w)$) along with 
$W_{141,1}(2)$ are shown.
\label{supfig:cmp}}
\end{figure}

\begin{figure}[ht]
\begin{minipage}[b]{0.45\linewidth}
\centering
\includegraphics[width=\textwidth]{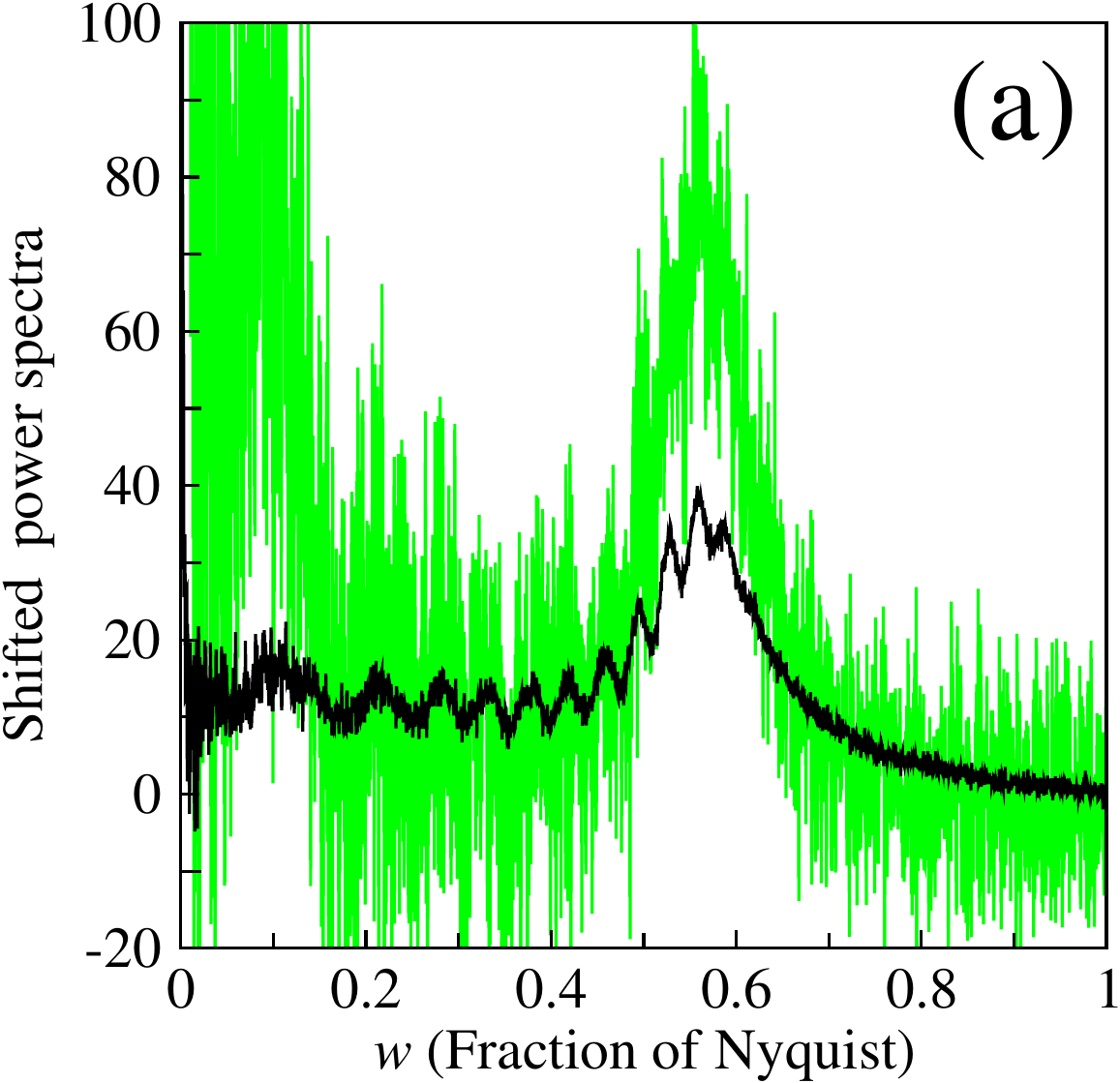}
\end{minipage}
\qquad
\begin{minipage}[b]{0.45\linewidth}
\centering
\includegraphics[width=\textwidth]{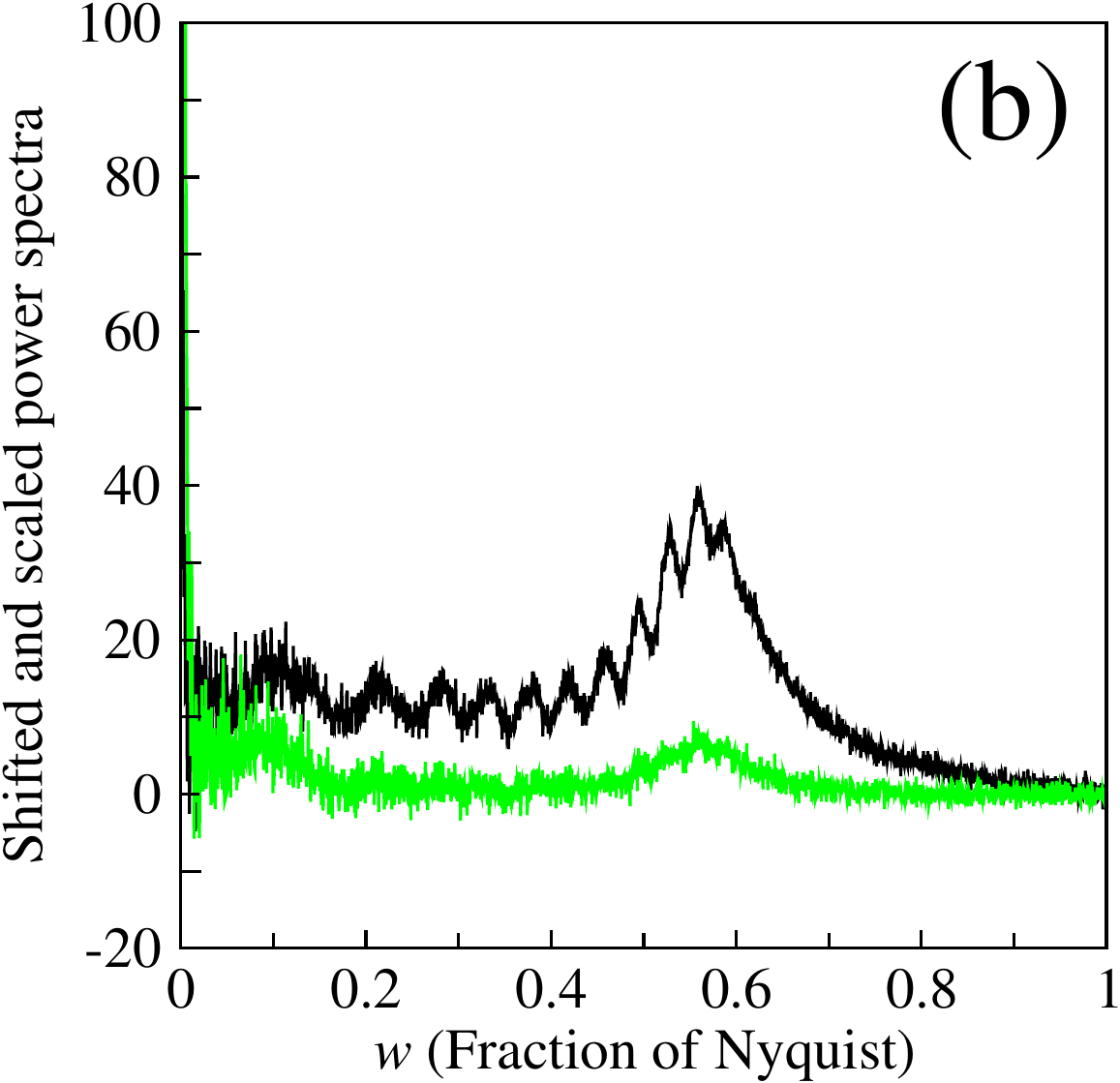}
\end{minipage}
\caption{Comparison of the circularly averaged Thon ring signal from the 
dose fractionated exposure (image \#230804) of amorphous ice  as seen
in the sum of the noise whitened power spectra (black) with that 
in the power spectrum of the sum of the image frames (green).  
The background offset, seen in 
Fig.~\ref{supfig:cmp}(b), has been removed by subtracting the average 
value near the Nyquist frequency.  The results from 
Fig.~\ref{supfig:cmp}(b) shifted in this way are 
shown in (a) while in (b)
the visibility of the Thon ring signal is shown by rescaling the power spectrum of the 
sum of the images in (a) so that it has the same noise level as sum of the power spectra curve.
\label{supfig:DEF}}
\end{figure}

\section{Comparison of $S(u)$ and $W(u)$ from exposure \#230804 of Fig.\ref{fig:ICE_THON} }
\label{sup:compare}
The effect of noise whitening is illustrated in Fig.~\ref{supfig:cmp}(a) which compares 
the spatial frequency dependence of the  sum of the circularly averaged power spectrum, $S_{141,1}$ 
and noise whitened power spectrum, $W_{141,1}$, (using the notation of Eq.~\eqref{eqn:WM}) 
obtained from the 141 frames used to generate Fig.~1 of the text.
The variation with spatial frequency of $S_{141,1}(w)$ is essentially that of 
${\mathcal N}(w)$. The multiplicative factor, $\Gamma$, relating $S_{141,1}(w)$  and
${\mathcal N}(w)$ is given in Eq.~\eqref{eqn:WX}. The measured value for $\Gamma$ of 684 is 
within 5\% of that from Eq.~(11) with a DQE(0) of ${\sim}$0.5 and dose of 2.33 e$^{-}$/pixel.
The noise whitening produced by division 
of $S(w)$ by ${\mathcal N}(w)$ removes this variation, producing a flat background above which the
amorphous ice Thon ring signal around 1/3.7\AA\/,  now enhanced by $1/{\mathcal N}(w)\sim 24$,  
can clearly be seen.  Noise whitening can be applied to both
the sum of the power spectra as shown in \ref{supfig:cmp}(a) and the power spectrum of the sum of the frames
given in \ref{supfig:cmp}(b). The lower noise in the sum of the power spectra provides a more stringent test
for the noise whitening procedure; the success of this is clearly illustrated by the featureless flat background 
in Figs. \ref{fig:ICE_THON} -- \ref{fig:CARBON_RANDOM_THON} of the main text.  

Residual correlations in the detector, in particular from
the applied linear pixel gain correction, lead to a difference in the value for the sum of the
power spectra versus the power spectrum of the image sum. This is illustrated in Fig.~\ref{supfig:cmp}(b)
where the value of $W_{141,141}(w)$ is 5\% greater than that of $W_{141,1}$.

Fig.~\ref{supfig:DEF}(a) compares the spatial frequency dependence of the circularly averaged
noise whitened sum of power spectra and that of noise whitened power spectrum of the sum of the frames.  
For convenience, an offset equal to Nyquist frequency value has been subtracted.
The Thon ring signal is bigger in the power spectrum of the sum of the frames but 
the background noise is also bigger (by a factor of $\sqrt{141} \sim 12$). 
In Fig.~1 of the main text, the Thon ring signal from
the amorphous ice appears stronger in the sum of the power spectra (Fig.~1(b)) rather than
in the power spectrum of the sum of the images (Fig.~1(a)). This is because the images 
in Fig.~1 have been scaled so that background noise has the same amplitude. This effect of this
is illustrated in Fig.~\ref{supfig:DEF}(b) and is obtained by dividing the circularly averaged
power spectrum of the sum frames by $\sqrt{141}$ so that the noise level in the two
plots (green and black) is the same.

\clearpage

\begin{figure}[ht]
\centerline{\includegraphics[width=0.95\textwidth]{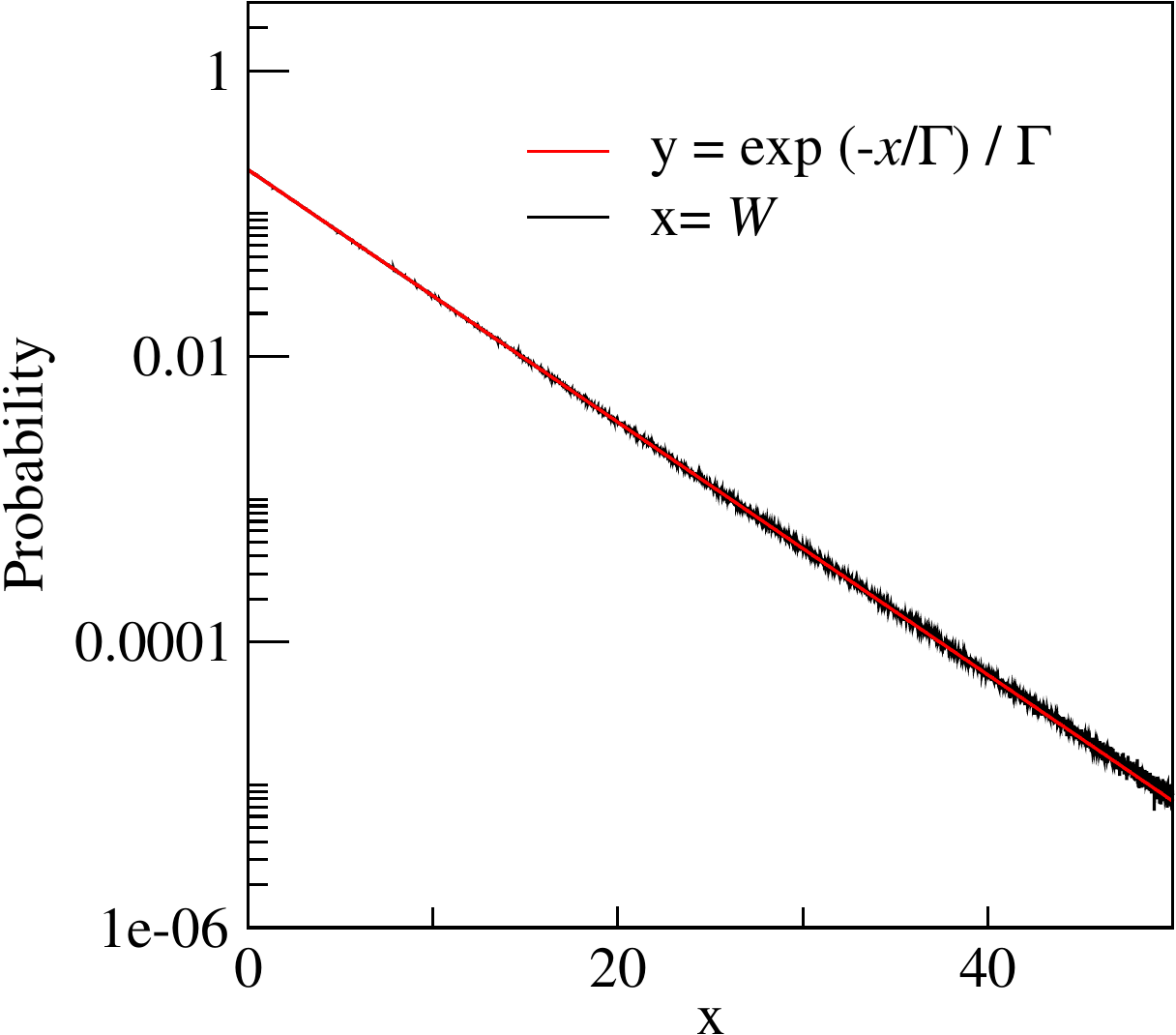}}
\caption{Histogram of the 141 frames noise whitened power spectra from 
image \#230804 of amorphous ice used in Fig.~1. The exponential 
distribution fit, {\color{red}--}, to 
the measured histogram, ---, is with $\Gamma= 4.91$.
\label{supfig:pwrhist}}
\end{figure}

\section{Statistics of noise whitened power spectra}
\label{sup:hist}
As discussed in the text, the distribution of the of values in a noise whitened power spectra 
of the Falcon~II is completely described by its mean value
$\Gamma \sim n/\mathrm{DQE}(0)$ in which $n$ is the 
mean number of incident electrons per frame, through 
exponential distribution of Eq.~(8). To illustrate this,
the histogram of values in 
noise whitened power spectra  of
the  141 frames 
used in Fig.~1 is given in
Fig.~\ref{supfig:pwrhist}. The measured histogram is an excellent fit
to an exponential distribution with parameter $\Gamma= 4.91$ (which 
differs by 5\% from the estimate based on $\mathrm{DQE}(0) = 0.5 $ and
$n =2.33\,$e$^{-}$/pixel/frame).

\clearpage

\begin{figure}[th]
\centerline{\includegraphics[width=0.90\textwidth]{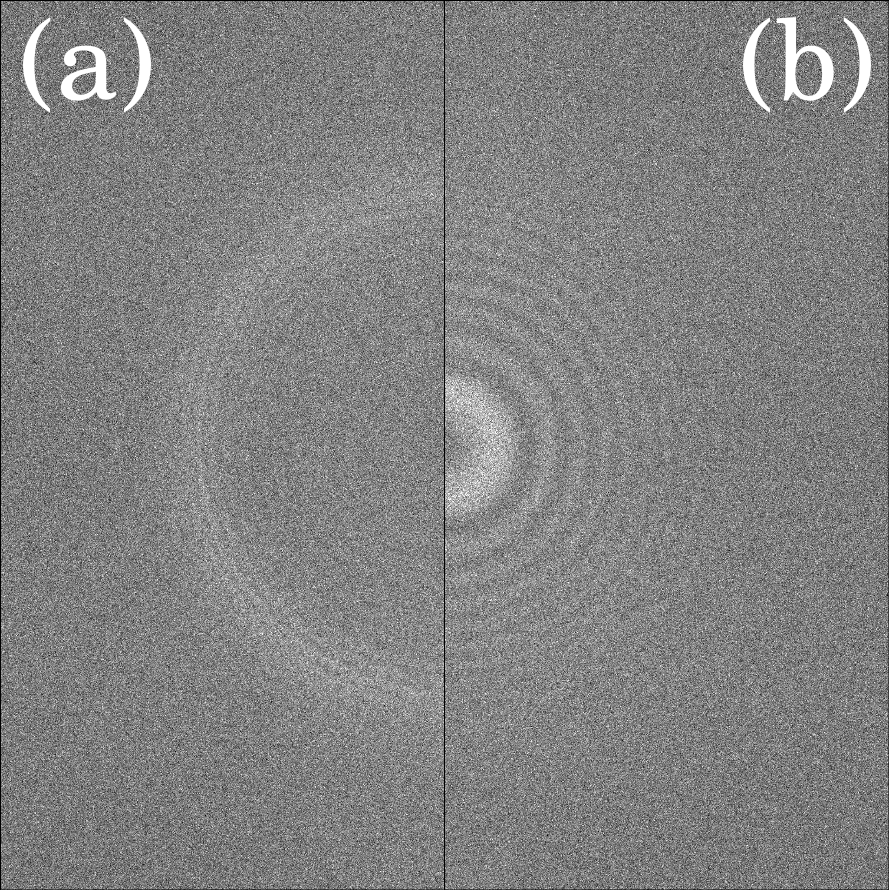}}
\caption{
(a) Sum of noise whitened power spectra from
the first 12 frames of image \#230804 from Fig.~1.
(b) Sum of noise whitened power spectra from
the first 12 frames of the pre-irradiated carbon image used in Fig.~2.
The same magnification and total dose  
($25.9 \sim 12\times 2.33/1.04^2\,$e$^{-}$/\AA$^2$) were used in both (a) and (b).
The Thon rings from amorphous ice around $1/3.7\,$\AA\/  are stronger 
than those from the pre-irradiated carbon sample.
\label{supfig:CMP_CARBON_ICE}}
\end{figure}

\section{Comparison at 26\,e$^{-}$/\AA$^2$ of Thon rings from amorphous ice with those 
from pre-irradiated amorphous carbon}
\label{sup:carbon_versus_ice}
\begin{figure}[htb]
\begin{minipage}[b]{0.45\linewidth}
\centering
\includegraphics[width=\textwidth]{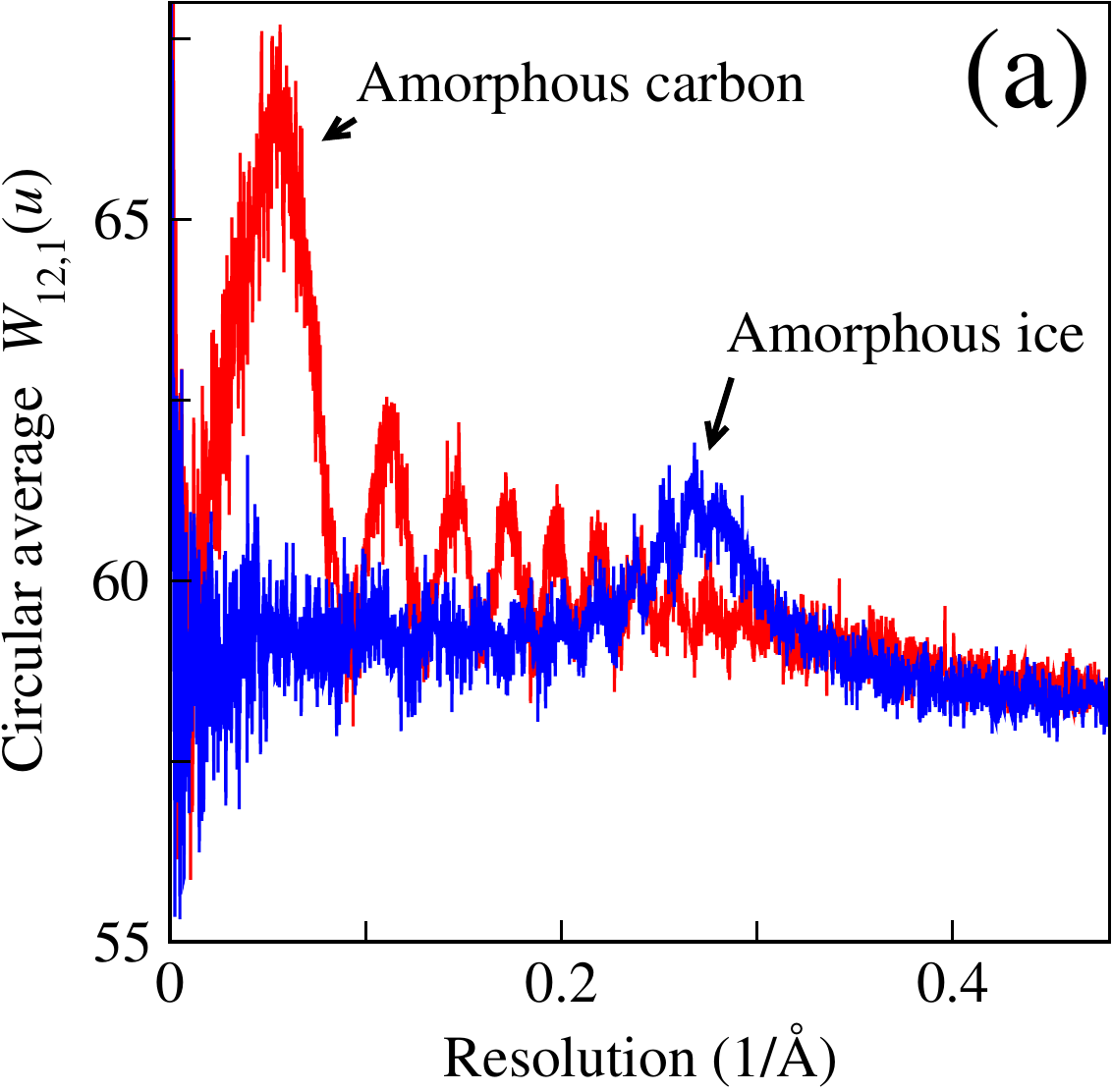}
\end{minipage}
\qquad
\begin{minipage}[b]{0.45\linewidth}
\centering
\includegraphics[width=\textwidth]{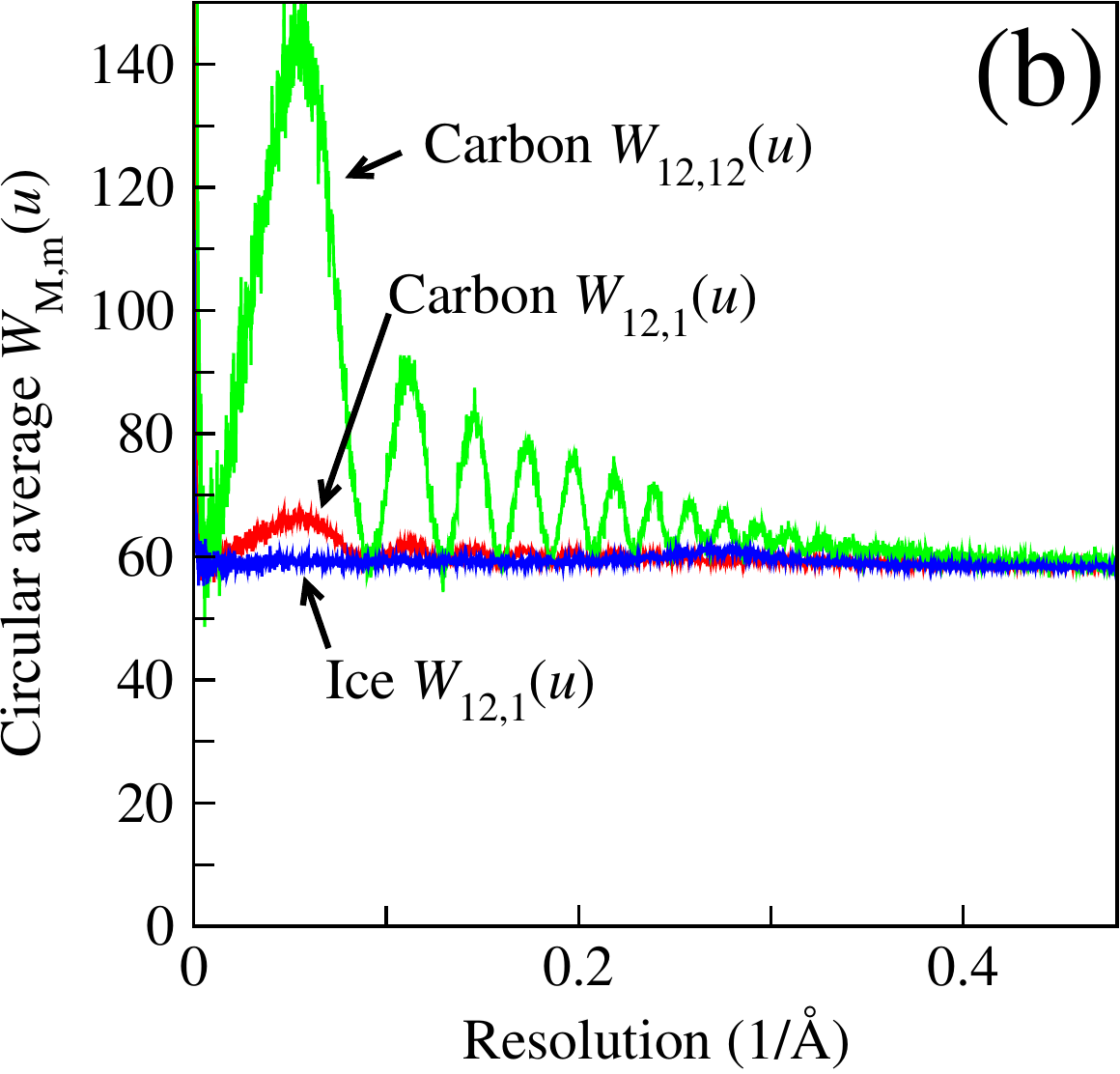}
\end{minipage}
\caption{
Comparison of the circular averages of the noise whitened power spectra obtained with 
$\sim 26 $e$^{-}$/\AA$^2$ from amorphous ice and pre-irradiated carbon. The $W_{12,1}(u)$ 
circular averages of the amorphous ice (blue) and pre-irradiated carbon (red) sums of
noise whitened power spectra shown in (a). In (b) the corresponding circular average
of the power spectrum, $W_{12,12}(u)$,  from the sum of the 12 frames of pre-irradiated amorphous 
carbon is shown.
\label{supfig:CIRC_CMP_CARBON_ICE}}
\end{figure}

Fig.~\ref{supfig:CMP_CARBON_ICE}(a) shows the Thon rings obtained from the 
sum of the first 12 noise-whitened power spectra, corresponding to 25.9\,e$^{-}$/\AA$^2$ from the 
141 used in Fig.~1.  For comparison Fig.~\ref{supfig:CMP_CARBON_ICE}(b) shows the corresponding
sum of noise whitened power spectra from images of pre-irradiated carbon in Fig.~2.
While the Thon rings
from carbon are stronger at low spatial frequency, around 1/3.7\AA\/ those from 
the amorphous ice sample are stronger. This is more clearly illustrated in the circular averages
shown in Fig.~\ref{supfig:CIRC_CMP_CARBON_ICE}.

Fig.~\ref{supfig:CIRC_CMP_CARBON_ICE}(a) shows the circular averages of the sum of
the noise whitened power spectra shown in Fig.~\ref{supfig:CMP_CARBON_ICE}.
In Fig.~\ref{supfig:CIRC_CMP_CARBON_ICE}(b) the
noise whitened power spectrum of the sum of the 12 pre-irradiated carbon frames
used in Fig.~\ref{supfig:CIRC_CMP_CARBON_ICE} is compared with 
the sum of 12 power spectra given in Fig.~\ref{supfig:CIRC_CMP_CARBON_ICE}(a). 
As the pre-irradiated carbon does not move during an exposure the Thon ring signal
adds coherently with increasing number of frames. The Thon ring modulation 
in the sum of the images, $W_{12,12}(u)$,  is 12 times that in the corresponding 
sum of power spectra, $W_{12,1}(u)$.
At the Nyquist frequency the image consists almost entirely of noise
and the power spectra of the sum of the images is equal to the 
sum of the power spectra. The noise in the power spectrum of the sum is however
$\sqrt{12}$ times that in the sum of the power spectra.

\clearpage
\section{Calculation of normalised noise power spectrum,  ${\mathcal N}(u) $}
\label{sup:nu}
The energy lost by incident high energy electrons passing through 
the lightly doped semi-conducting sensitive layer of a CMOS sensor such as the Falcon~II, produces
electron/hole pair excitations. The sensitive layer is surround by heavily P-doped layers into
which the holes are free to escape but the negatively charged electron excitations 
are trapped due to the potential resulting from the differential doping. 
An incident high energy electron is recorded via the voltage drop on 
reverse biased diodes formed by a heavily
N-doped implants on the surface of the sensitive layer resulting for
collection of the electron excitations. 
The number of electron-hole pairs
generated by an incident high energy electron, and hence the signal, depends on the length 
of the incident electrons trajectory through the sensitive layer. The thick sensitive
layer of the Falcon~II results in a large average signal but the thickness of the 
sensitive layer allows electrons to diffuse relatively large distances before they are 
finally collected.  This diffusion results in
a poor modulation transfer function, MTF, even with the relatively large
 ($14\,\mu$m) pixel size of the Falcon~II.
The diffusion acts as a low-pass filter and while it lowers the MTF it has only
a small effect on the detective quantum efficiency, DQE. In other words the 
spatial distribution of the signal deposited in the detector by an incident
electron, described by the point spread function (PSF), is mainly determined 
by electron/hole diffusion and
not scattering of the primary incident electron (at least for incident 300 keV electrons). 
The Fourier transform of the point spread
function determines the intrinsic spatial frequency response of the detector and 
its square describes the noise power spectrum. 
In a real pixellated detector the noise power spectrum
also includes pixel modulation terms of the form $\sin(x)/x$ from integration over the 
pixel and contributions aliased back from 
spatial frequencies beyond the Nyquist frequency
of the pixellated sampling. 

As shown in \cite{mcmullan_comparison_2014}   the 
${\mathcal N}(u)$ for the Falcon~II detector is to a good approximation given by
\begin{equation}
{\mathcal N}(u,v) = \sum_{\bf G} 
\bigg( \sin (\pi a u_G )\sin( \pi a v_G)  \, \mathrm{MTF}_0( \sqrt{ u_G^2 + v_G^2} )\bigg)^2
\label{eqn:supp_Nu}
\end{equation}
where $a$ is the pixel pitch, the sum is over the low order reciprocal lattice vectors
needed to describe the aliasing, $u_G$ and $v_G$ are the spatial frequency components i.e., 
$u_G = u + G_u$, and $\mathrm{MTF}_0$ is the Fourier transform of the intrinsic PSF.
In \cite{mcmullan_comparison_2014}   ${\mathcal N}(u,v)$ was calculated from the 
$\mathrm{MTF}_0$ obtained from the experimentally measured MTF. While
this works well (reducing the variation in the power spectra to a few percent), 
the higher signal to noise in the longer exposures used
in the present work show small systematic variations. The results presented
in Figs. 1-3 were therefore obtained using a fit to a radial function $\phi(u)$, effectively $\mathrm{MTF}_0(u)^2$,
from zero to twice the Nyquist frequency of the measured noise power spectra
using the functional form in Eq.~\eqref{eqn:supp_Nu}. 

The simplest way to calculate  ${\mathcal N}(u,v)$ is to average the noise power spectra
of many frames and apply the symmetry from the square pixel array. This leaves some 
residual noise and the problem of finding the overall normalisation. In the present work
a radial function, $\phi(u)$, was 
used to fit the symmetrized sum of the noise power spectra via
\begin{equation}
{S}(u,v) \sim  \sum_{\bf G} 
\sin^2 (\pi a u_G )\sin^2( \pi a v_G)  \, \phi( \sqrt{ u_G^2 + v_G^2} ).
\label{eqn:supp_fit}
\end{equation}
The overall normalisation was found by using a single Gaussian function in $\phi(u)$ to
fit the low spatial frequency limit and a smooth general fit to the 
remainder via a b-spline fit that was penalty weighted to ensure smoothness 
and non-negative values.  
A more detailed description of this and its implementation will be presented elsewhere.

\end{document}